\documentclass[%
 aip,
 amsmath,amssymb,
 reprint,%
]{revtex4}

\usepackage{graphicx}
\usepackage{dcolumn}
\usepackage{bm}

\usepackage[utf8]{inputenc}
\usepackage[T1]{fontenc}
\usepackage{mathptmx}

\begin{document}

\preprint{AIP/Physics of Fluids}

\title{Two typical collective behaviors of the heavy ions expanding in cold
plasma with ambient magnetic field}

\author{Guo-Liang Peng}
\affiliation{Beijing Institute of Technology, Beijing 100081, China}
\affiliation{Northwest Institute of Nuclear Technology, Xi'an 710024, China}

\author{Jun-Jie Zhang}
\affiliation{Northwest Institute of Nuclear Technology, Xi'an 710024, China}

\author{Jian-Nan Chen}
\affiliation{Northwest Institute of Nuclear Technology, Xi'an 710024, China}

\author{Tai-Jiao Du}
\affiliation{Northwest Institute of Nuclear Technology, Xi'an 710024, China}

\author{Hai-Yan Xie}
\affiliation{Northwest Institute of Nuclear Technology, Xi'an 710024, China}

\date{\today}

\begin{abstract}
We have numerically studied the evolution of the heavy ions that expand
in a cold background plasma at a large scale. Two typical collective
behaviors of the heavy ions are identified with the conditions where
only the traversing heavy ion's initial total mass is different. Our
work has demonstrated that a difference in the initial total mass
of the moving heavy ions is able to induce completely different collective
behaviors of the plasma. The simulation is performed via the hybrid
model, in which the ions and electrons are treated as classical particles
and mass-less fluid, respectively. Due to the imbalance of the electric
and magnetic force on the heavy ions, these particles will evolve
into different collective patterns at the later time. These patterns
manifest a rather different stopping behavior of the moving ions and
an opposite drifting direction of the electron fluid at the rim of
the expanding plasma. Further numerical and analytical calculations
show that the imbalance depends not only on the number densities of
the plasma ions, but also on the spatial variations of the magnetic
fields. Our work reveals that the collective behavior of the heavy
ions is highly non-linear, and the non-linearity is able to induce
different phenomena in the evolution of the system at a large scale.
\end{abstract}

\maketitle

\section{Introduction}

When a patch of heavy ions expands and traverses a magnetized background
plasma with a high velocity, a super-Alfvénic shock will be formed
at the front of the moving ions\cite{HEWETT2011,Brecht2009}. In such
a condition, the heavy ions disturb the background plasma and may
form a magnetic cavity\cite{winske2019} at a large scale. Our present
work has identified two typical collective behaviors of the heavy
ions expanding in a dilute cold background plasma with the existence
of an ambient magnetic field. These conditions have many practical
and potential correspondences, such as the solar wind in near space
environment\cite{Pognan2018,Goldstein2005,Hofmeister2020}, planetary
atmospheric plasma\cite{Kim2020,Yamazaki2020,Bultel2012,Campbell2013}
and other similar effects\cite{Zinn1963,Raizer1995,McDoniel2019,Sinibaldi2019,Xu2019}.

In these conditions, the mean-free-path of the expanding heavy ions
usually takes a few hundred kilometers. As the ions (e.g. iron ions)
expand in the dilute background plasma, collisionless shocks\cite{Treumann2009,Ryutovl2018}
will appear and influence the later collective behavior of these ions.
Along with the ion-evolution, a flute mode (caused by the flute instability\cite{Akimune1981})
may be generated before they finally stop moving\cite{Akimune1981, KOPECKY1968}.

Due to the non-linearity of the plasma and the limited experimental
data at such a large scale (a few hundred kilometers), various phenomena
of this kind of physical process remains to be investigated. This
paper focuses on the stopping behavior and mass influence of the heavy
ions. The simulation is performed via the hybrid model\cite{Garyand1990,HARNED1982,HEWETT1980,HEWETT2011,Thomas1986,Thomas1987,Winske2007},
which treats the ions as PIC (Particle-In-Cell) particles and the
electrons as mass-less fluid to neutralize the plasma. To understand
how the initial mass of the heavy ions influences their subsequent
behavior, we have chosen two typical conditions where only the initial
mass of the heavy ions is different. These two conditions correspond
to the atmospheric phenomena in near space\cite{Artemyev2019,Goldstein2005,Hofmeister2020,Pognan2018}
($\Gamma\ll1$, expression of $\Gamma$ is in Eq. (\ref{eq:gamma}))
and the laser experiments ($\Gamma\geq1$)\cite{Howes2018,Sinibaldi2019,Shaikhislamova2015,Schaeffer2017,Ganguli2015,Valenzuela1986,Heuer2018}. 

Our studies find that if the total mass of the heavy ions are small,
they rotate in the ambient geomagnetic field and couple weakly with
the background plasma (Fig. \ref{fig:Motion-of-the}), which corresponds
to the decoupled phenomena in Ref. \cite{HEWETT2011}. The collective
rotation of these ions produces a breathing pattern (Fig. \ref{fig:Evolution-of-the-1}
and \ref{fig:Evolution-of-the-2}) which is quite different from the
$\Gamma\ll1$ simulation (Fig. \ref{fig:Evolution-of-the}), where
the ions expand nearly in a straight line and curve at the end without
a breathing pattern (Fig. \ref{fig:Motion-of-the-1}). To analyze
the possible mechanism for the breathing pattern, we have chosen four
heavy particles, and depicted their electric and magnetic forces (Fig.
\ref{fig:Motion-of-the} and \ref{fig:Motion-of-the-1}). From the
analysis of the forces, we find that the imbalance of the electric
and magnetic forces in the $\theta$ direction is the main contributor
to the breathing pattern and the flute mode (Eq. (\ref{eq:u_b u_d})).
Moreover, the directions of the electron drift are also different
at the rim of the expanding torus in the two cases (Fig. \ref{fig:Directions-of-the}).
The difference of the electron drift can affect the stopping behavior
of the ions (Fig. \ref{fig:Trajectories-of-the}).

The paper is organized in the following structure. In section \ref{sec:The-simulation-model},
we have briefly introduced the theoretical set-up of the hybrid model
and performed a $\Gamma\ll1$ simulation. In section \ref{sec:Two-typical-patterns},
we have presented the simulation results of the breathing pattern
and depicted the relevant forces and motion of the debris ions. The
conclusion is made in section \ref{sec:Summary-and-conclusion}. 

\section{The simulation model\label{sec:The-simulation-model}}

When the gyro-radii of the heavy ions are about a few hundred kilometers
which are comparable to the ion mean-free-path, the Knudsen number
is larger than one and the kinetic effect plays a more important role
than the fluid effect\cite{Artemyev2019}. In this condition, MHD
(Magneto-Hydro Dynamics) calculations cannot be trusted. In this work
we use the so called hybrid plasma simulation model to perform the
particle evolution and analyze the collective motion of the heavy
ions. 

The hybrid model starts with the equation of motion for electrons.
In S.I. unit this reads\cite{HEWETT2011}
\begin{eqnarray}
m_{e}n_{e}\frac{d\mathbf{u}_{e}}{dt} & = & \nabla(n_{e}kT_{e})+en_{e}(\mathbf{E}+\mathbf{u}_{e}\times\mathbf{B})-en_{e}\eta(\mathbf{J}_{i}+\mathbf{J}_{e}),\label{eq:EOM of electrons}
\end{eqnarray}
where $m_{e}$, $n_{e}$, $\mathbf{u}_{e}$, $T_{e}$ and $\mathbf{J}_{e}$
are the mass, number density, fluid velocity, temperature and electric
current for electrons, receptively. Vectors $\mathbf{E}$ and $\mathbf{B}$
denote the electric and magnetic field acted upon the electrons. Constants
$e$, $k$ and $\eta$ are the electronic charge, the Boltzmann constant
and the electric resistivity. $\mathbf{J}_{i}$ represents the ion
electric current. 

Currently we mainly focus on the conditions where the typical electron
gyro-radius and Debye length are about a few to a hundred centimeters.
In these conditions, the detailed motion of the electrons are negligible
compared to that of the ions. Thus, it is reasonably acceptable to
assume that: i) all electrons are mass-less and ii) the electric field
applied on the electrons are so strong that the electrons respond
instantly to the electric field and always neutralize the ions\cite{HEWETT1980}.
These assumptions reduce the computational cost needed for tracking
the information of electrons. For mass-less electrons, the left hand
side of Eq. (\ref{eq:EOM of electrons}) equals zero. For numerical
convenience, we may further assume that the gradient of the electron
temperature and the electric resistivity are small\cite{HEWETT2011},
hence, we can explicitly express $\mathbf{E}$ from Eq. (\ref{eq:EOM of electrons})
as,
\begin{eqnarray}
\mathbf{E} & = & -\mathbf{u}_{e}\times\mathbf{B}.\label{eq:E in B and U_e}
\end{eqnarray}

Since we focus on the ion-motion in large spatial and time scales,
the propagation of high-frequency radiations is not of much interest.
This allows us to drop the displacement current in Maxwell's equations
and simply adopt the Ampere's law for the electron fluid velocity
$\mathbf{u}_{e}$,
\begin{eqnarray}
\nabla\times\mathbf{B} & = & \mu_{0}(\sum_{\text{ion-species}}\mathbf{J}_{i}+\mathbf{J}_{e})\nonumber \\
 & = & \mu_{0}(\sum_{\text{ion-species}}\mathbf{J}_{i}-en_{e}\mathbf{u}_{e}),\label{eq:Ampere's law}
\end{eqnarray}
where $\mu_{0}$ is the vacuum permeability. Note that Eq. (\ref{eq:Ampere's law})
is a rather stronger approximation than the usual Darwin limit\cite{HEWETT1980,Gibbons1995}.
In Darwin limit only the solenoid part of the displacement current
is neglected. This means that in our calculation both $\mathbf{B}$
and $\mathbf{E}$ reach the asymptotic limit at each time step. Combining
Eq. (\ref{eq:E in B and U_e}) and (\ref{eq:Ampere's law}), we can
write the electric field $\mathbf{E}$ in terms of magnetic field
$\mathbf{B}$ and ion current $\sum_{\text{ion-species}}\mathbf{J}_{i}$,
\begin{eqnarray}
\mathbf{E} & = & -\frac{1}{en_{e}}\left(\sum_{\text{ion-species}}\mathbf{J}_{i}-\frac{\nabla\times\mathbf{B}}{\mu_{0}}\right)\times\mathbf{B}.\label{eq:E}
\end{eqnarray}
Finally we use 
\begin{eqnarray}
\nabla\times\mathbf{E} & = & -\frac{\partial\mathbf{B}}{\partial t}\label{eq:B}
\end{eqnarray}
and
\begin{eqnarray}
n_{e} & = & \sum_{\text{ion-species}}(Z_{i}n_{i})\label{eq:neutral condition}
\end{eqnarray}
to close Eq. (\ref{eq:E}), where $Z_{i}$ and $n_{i}$ denote the
charge state and number density of the ions, respectively. In Eq.
(\ref{eq:neutral condition}) we have used the above mentioned neutrality
assumption. 

To solve Eqs. (\ref{eq:E}) $\sim$ (\ref{eq:neutral condition}),
the finite difference method is utilized with all ions treated as
PIC particles. Given the initial distribution of the background plasma
and the rapidly moving heavy ions in phase space, the evolution of
all ions in the ambient magnetic field can be obtained accordingly.

\begin{figure}
\begin{centering}
\begin{tabular}{ccc}
\includegraphics[scale=0.24]{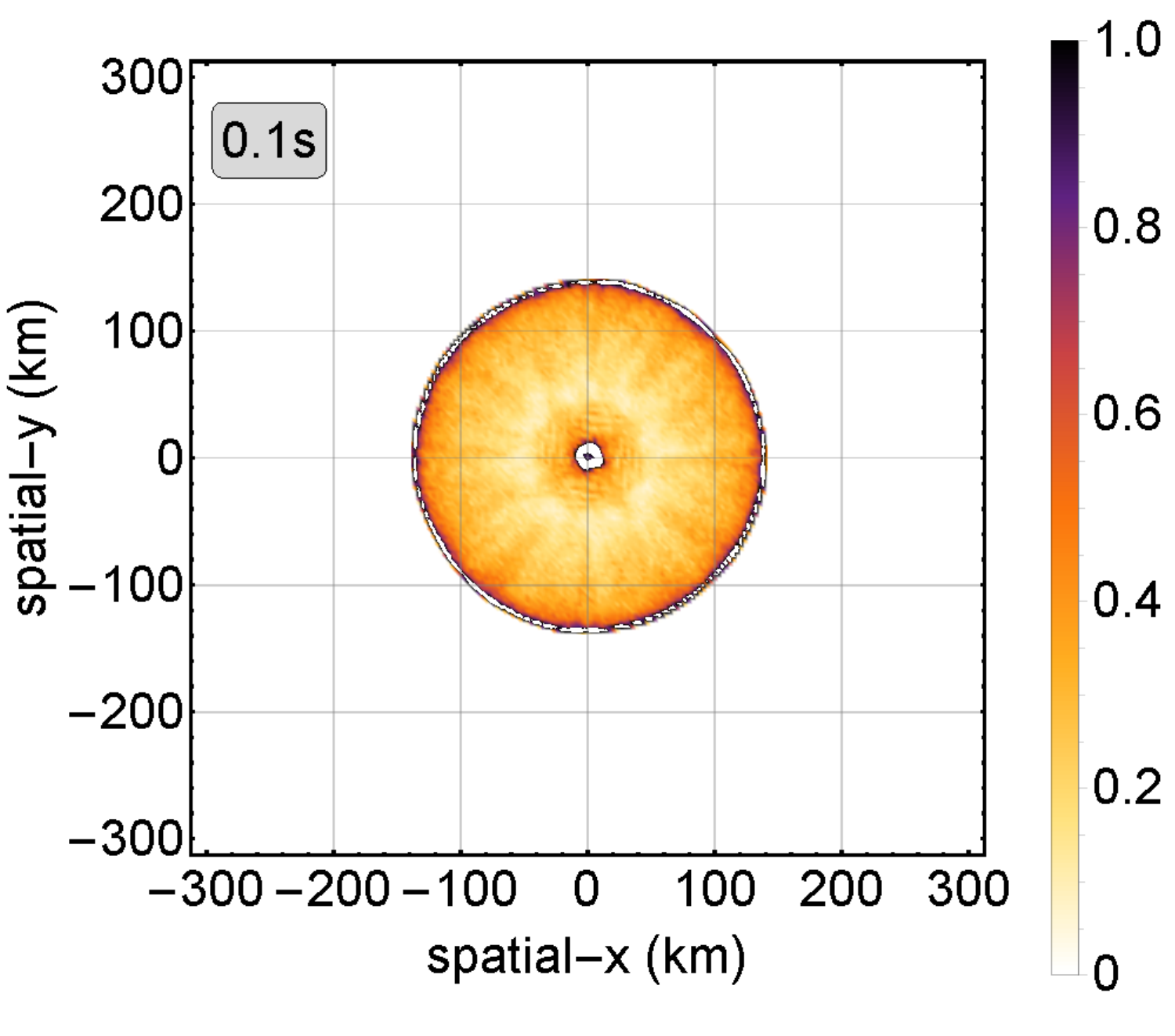} & \includegraphics[scale=0.24]{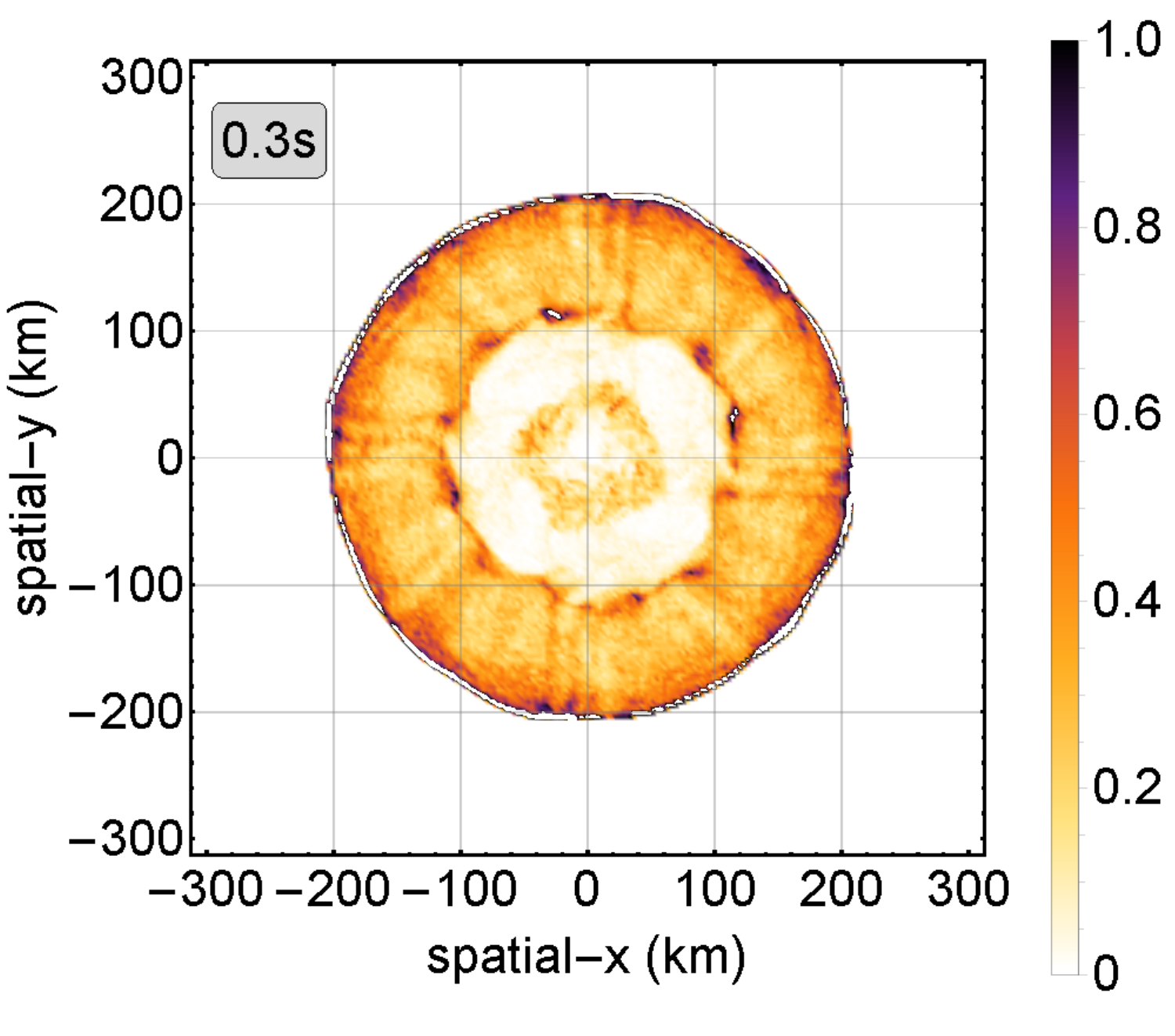} & \includegraphics[scale=0.24]{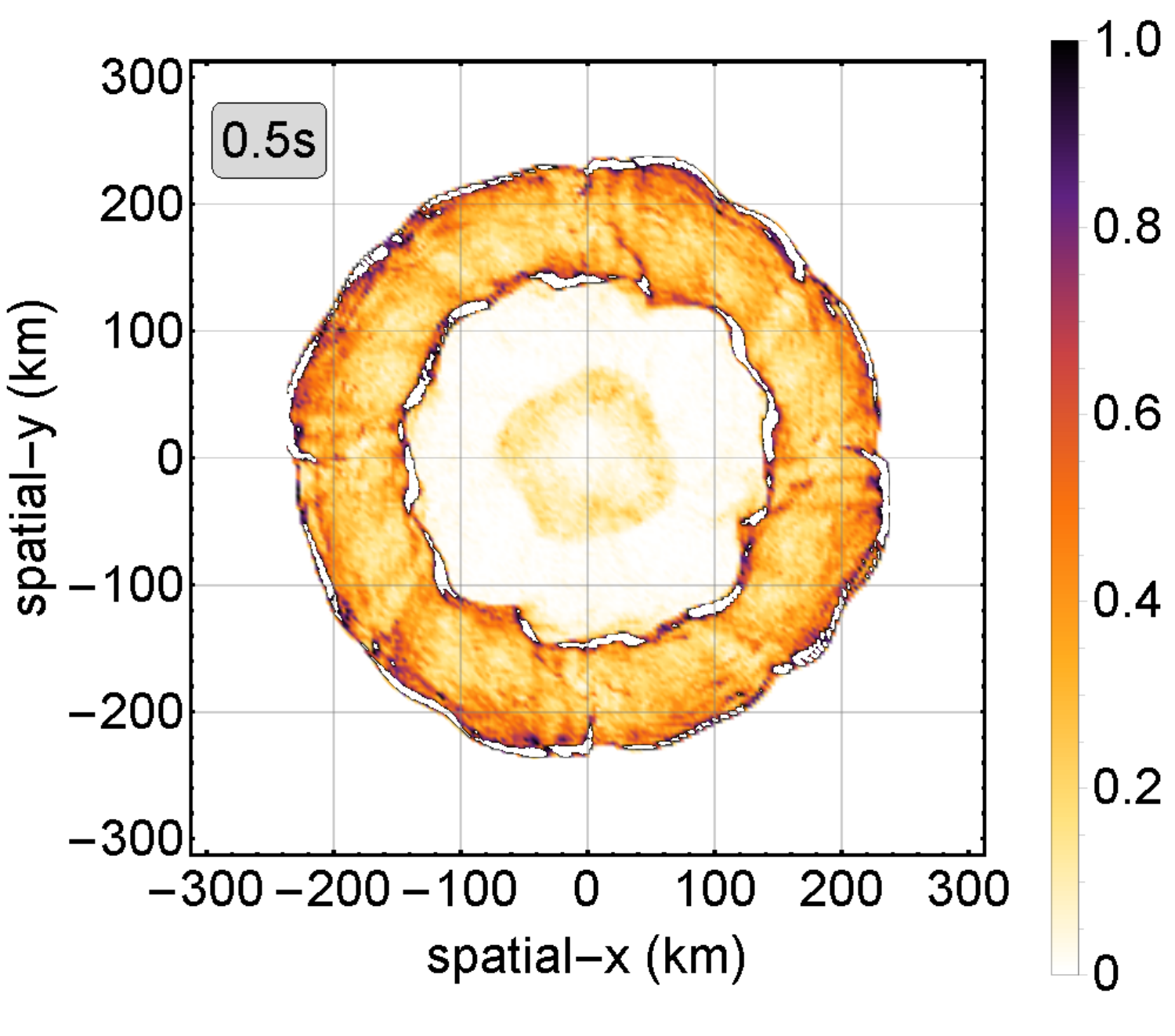}\tabularnewline
\includegraphics[scale=0.24]{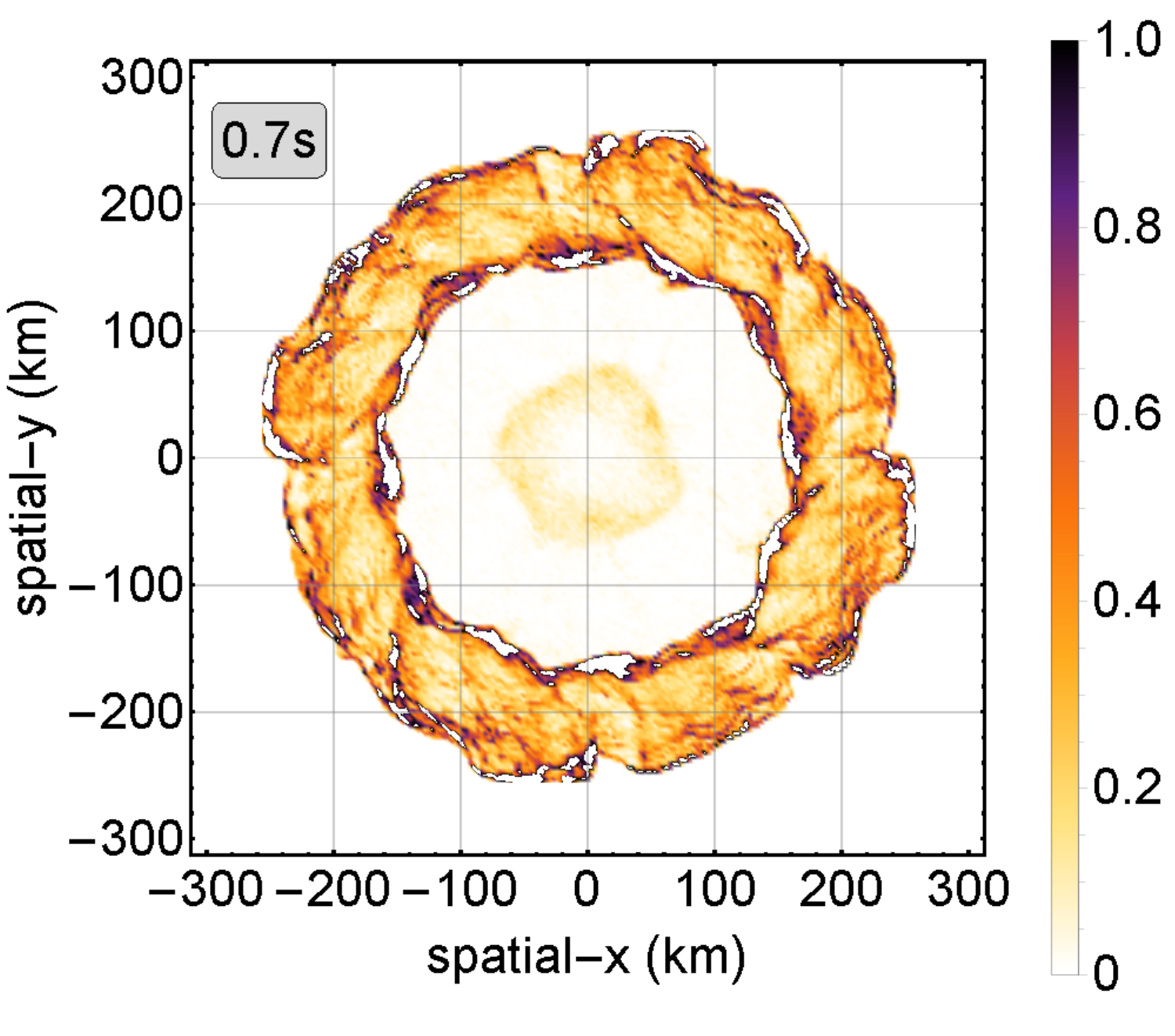} & \includegraphics[scale=0.24]{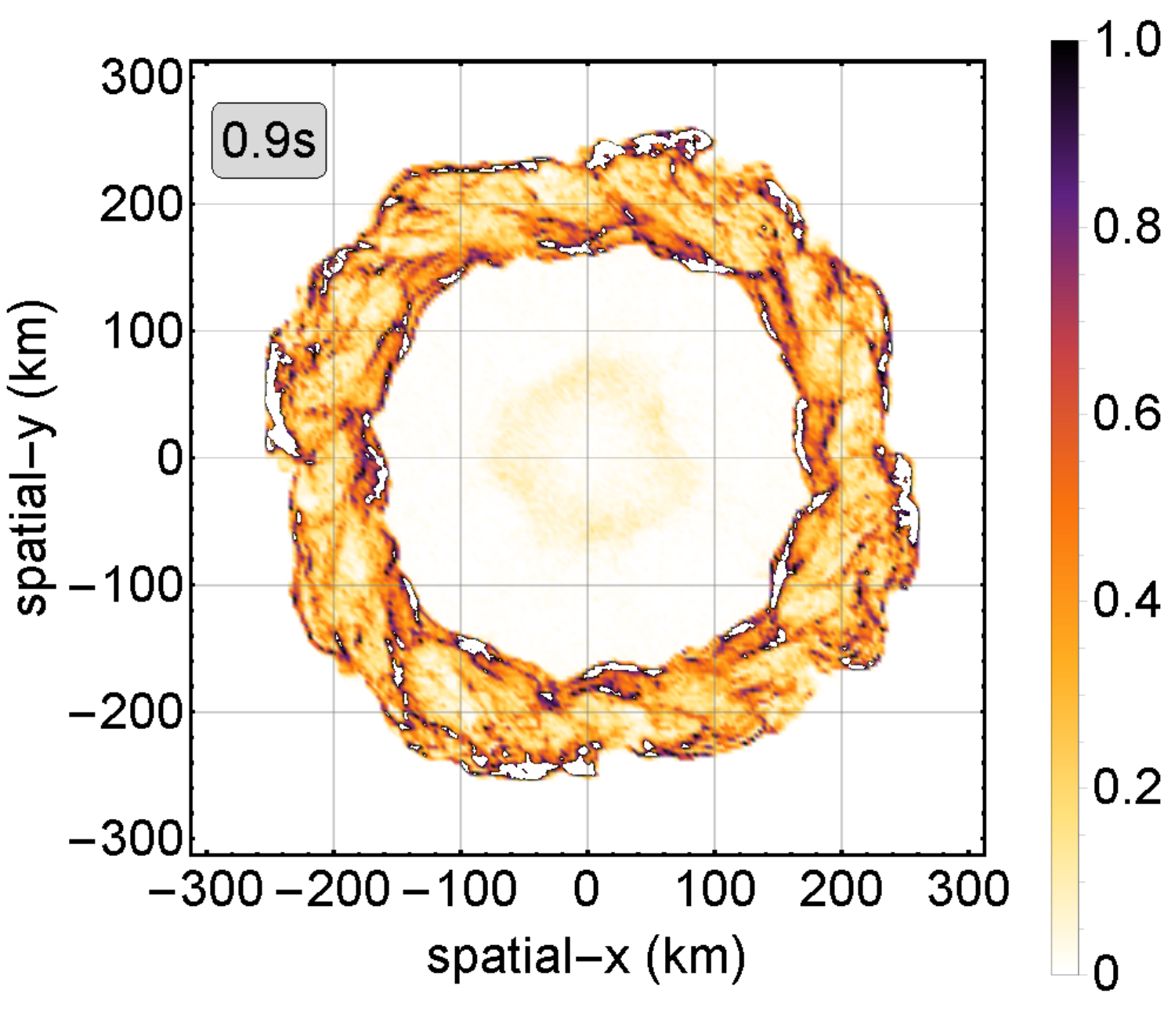} & \includegraphics[scale=0.24]{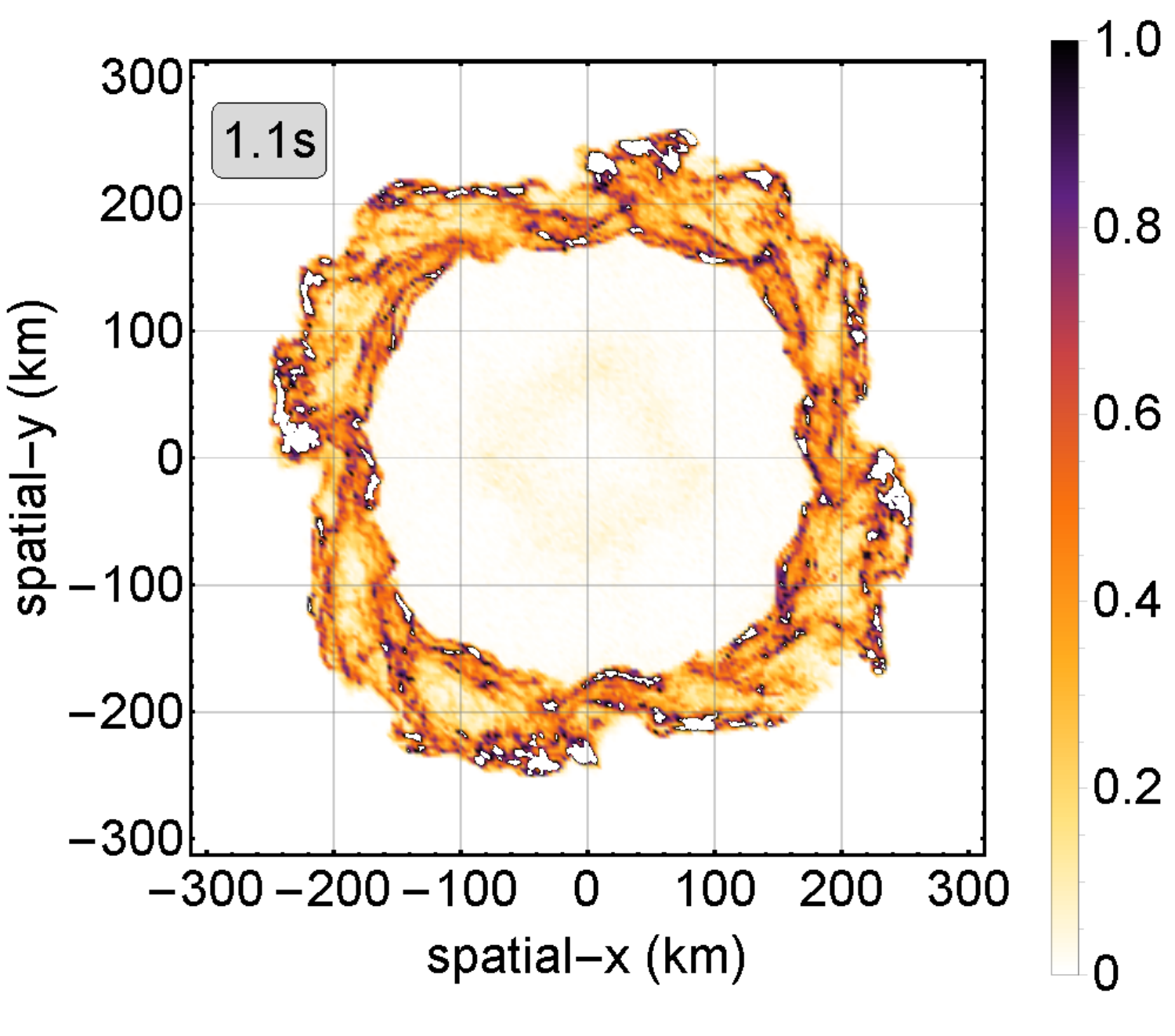}\tabularnewline
\end{tabular}\caption{Evolution of the heavy ions at various snapshots with $\Gamma\simeq0.03$.
The figures are viewed in the direction of the ambient magnetic field,
which points out of the plane with initial value $\mathbf{B}_{0}=0.5\times10^{-4}\ \text{T}$
throughout the spatial grids. The heavy ions are Iron particles with
charge state $Z_{i,\text{Fe}}=+1$. All heavy ions are initially placed
at spatial point $(0,0,0)$ with radial velocity $v_{i,\text{Fe}}=2\times10^{6}\ \text{m/s}$.
$10^{3}\ \text{kg}$ heavy ions are used in total. The number density
of the background ions (Oxygen) are taken to be $10^{7}\ \text{/cm}^{3}$
with charge state $Z_{i,\text{O}}=+1$. One sees clearly that the
flute mode occurs at some later time steps. \label{fig:Evolution-of-the}}
\par\end{centering}
\end{figure}

For an physical intuition of the ions evolution, we have performed
a 3D simulation in which a patch of heavy ions expanding in the cold
plasma with the presence of ambient geomagnetic field (see Fig. \ref{fig:Evolution-of-the}).
The background particles are taken to be the Oxygen ions which are
distributed uniformly in spatial grids with a constant number density.
The heavy ions are Iron ions which initially locate at the point $(0,0,0)$.
The heavy ions traverse the background plasma at a constant initial
velocity and evolve into a thin shell with flute modes at some later
time. 

The flute mode is a typical phenomenon in this condition. To measure
its occurrence, the usual parameter used is the ratio\cite{Schaeffer2017}
\begin{equation}
\Gamma=\rho_{i}/R_{3},\label{eq:gamma}
\end{equation}
where $\rho_{i}$ is the direct ion gyro-radius and $R_{3}$ is the
magnetic confinement radius. $R_{3}$ can be obtained by equating
the initial kinetic energy of the heavy ions with the equivalent magnetic
field energy in a volume of radius $R_{3}$,
\begin{eqnarray}
E_{i} & = & \frac{1}{2}\frac{\mathbf{B}^{2}}{\mu_{0}}\frac{4}{3}\pi R_{3}^{3},\label{eq:R_3}
\end{eqnarray}
where $E_{i}$ is the total initial kinetic energy of the heavy ions.
When $\Gamma\gtrsim1$, like in laser or space experiments \cite{Howes2018,Shaikhislamova2015,Schaeffer2017,Ganguli2015,Valenzuela1986,Heuer2018},
the flute mode will be less clear. When $\Gamma\ll1$, a strong flute
mode will occur. 

One also sees the reflected ions in the middle of the torus at $t=0.3\ \text{s}$
in Fig. \ref{fig:Evolution-of-the}. Due to the finite computing region,
these reflected ions gradually run out of the computing domain and
eventually fade away. The reflected ions play an important role in
the formation of the shock structure and influence the stability of
the plasma\cite{Marcowith2016,Treumann2009,Yoshihara1961}. 

\section{Two typical patterns of the heavy ions\label{sec:Two-typical-patterns}}

\begin{figure}
\begin{centering}
\begin{tabular}{cccc}
\includegraphics[scale=0.17]{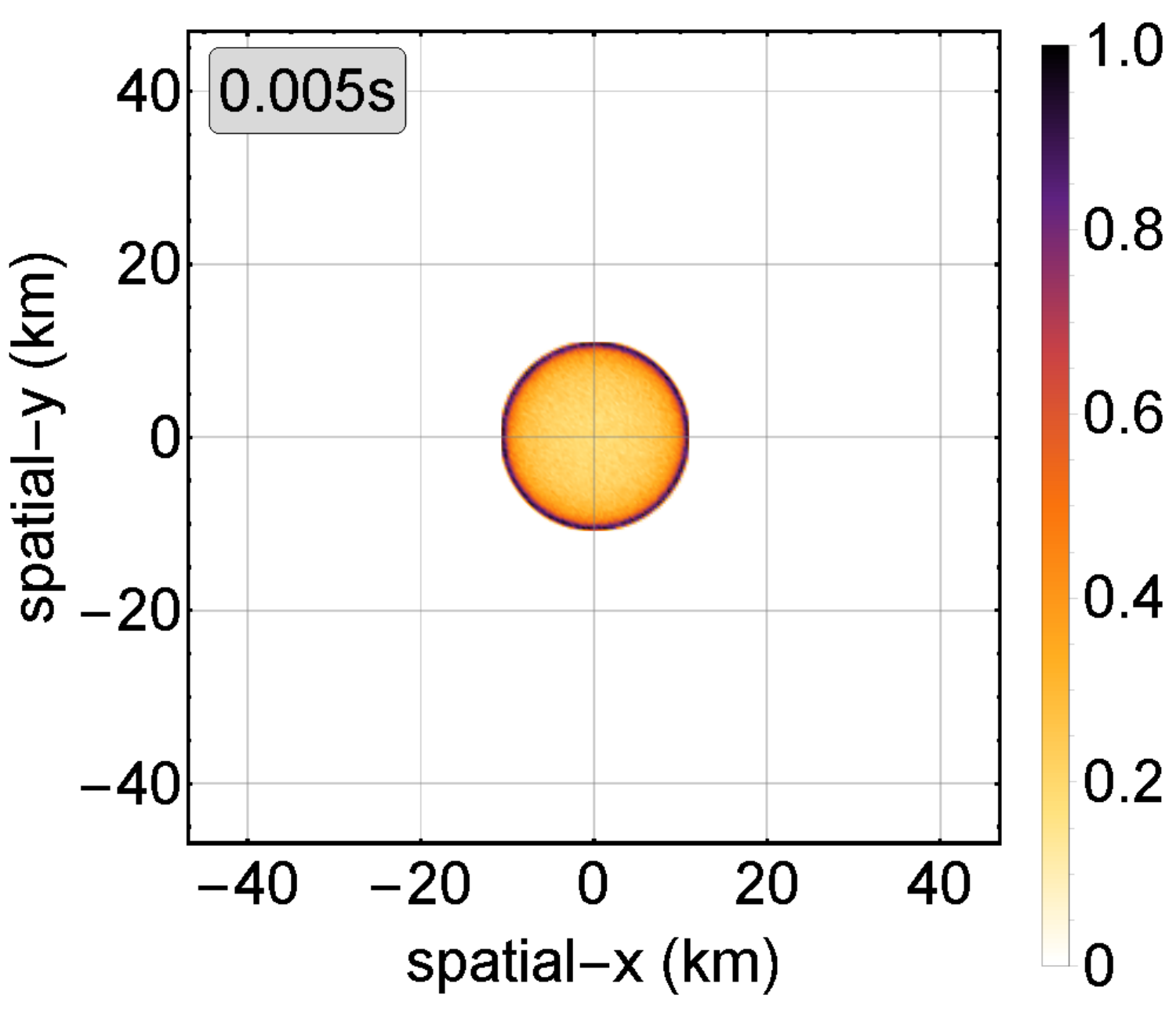} & \includegraphics[scale=0.17]{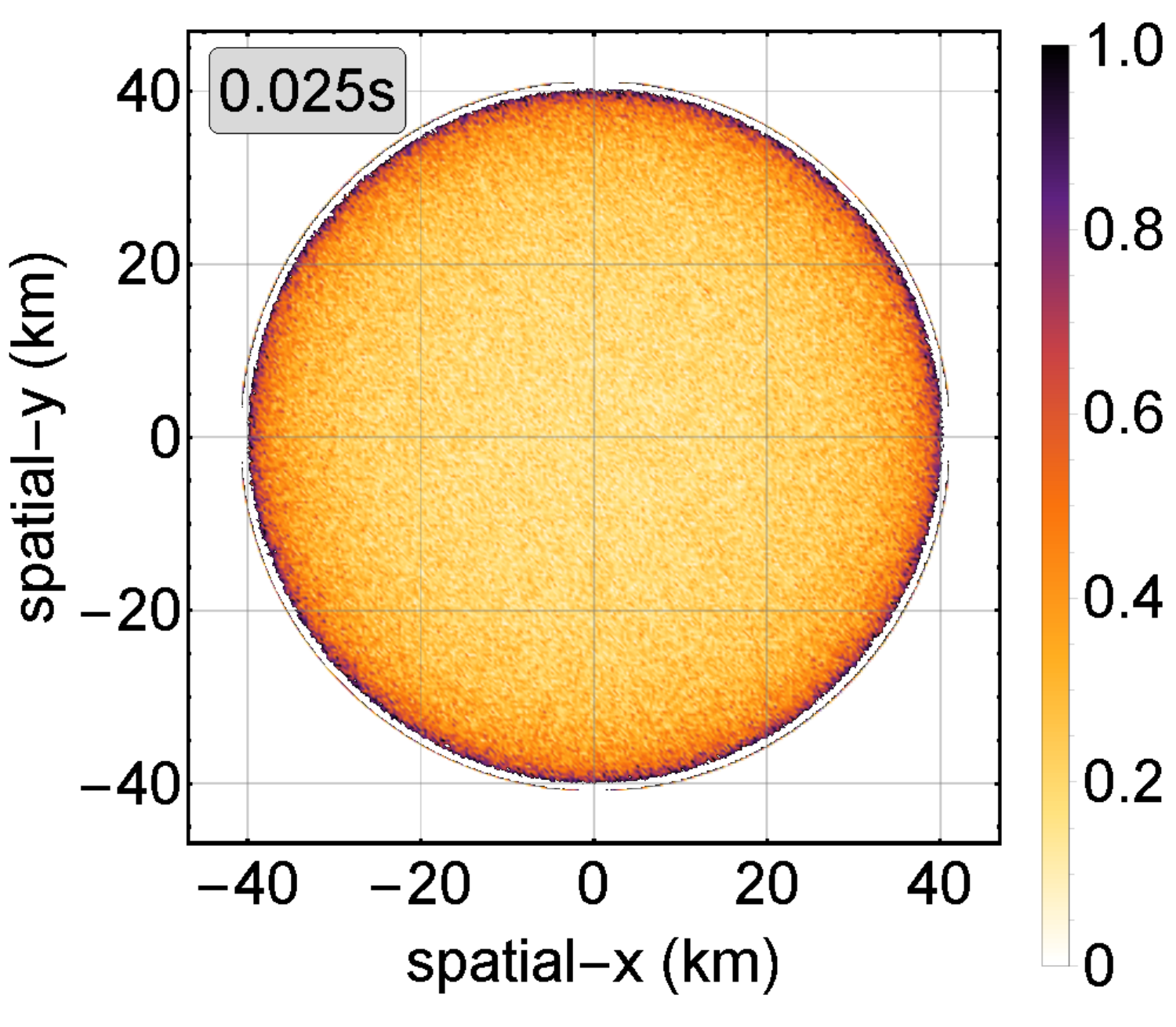} & \includegraphics[scale=0.17]{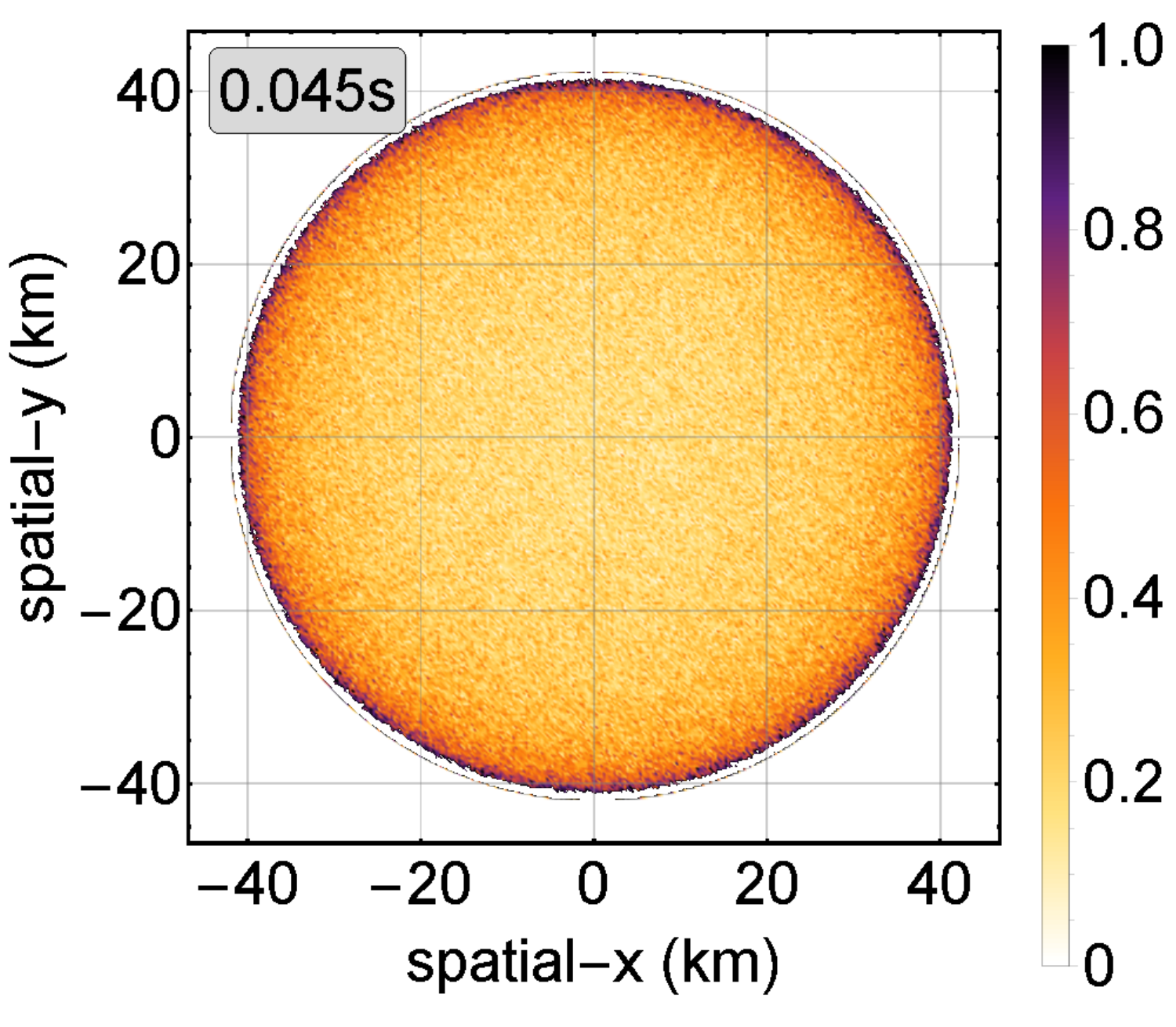} & \includegraphics[scale=0.17]{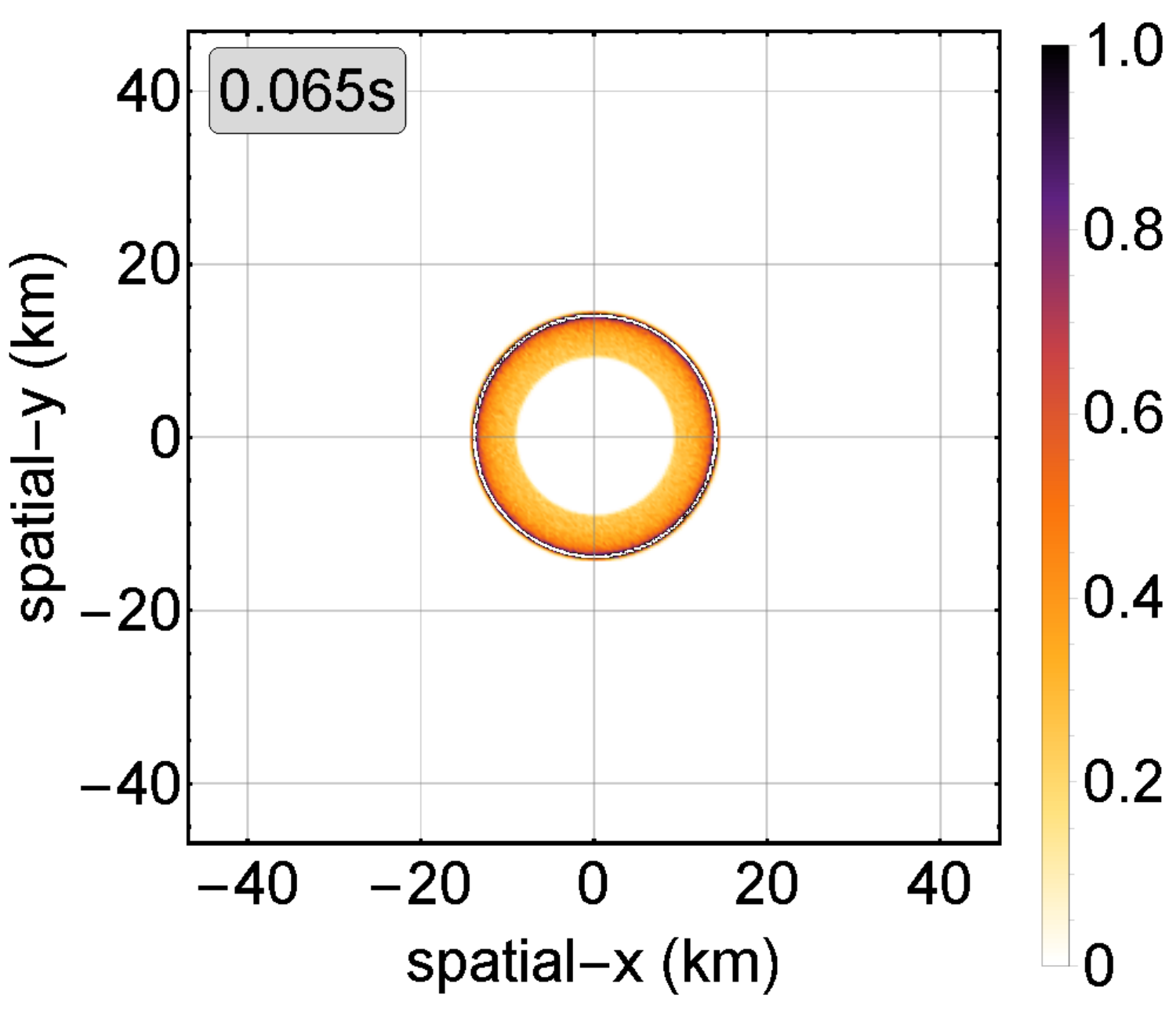}\tabularnewline
\includegraphics[scale=0.17]{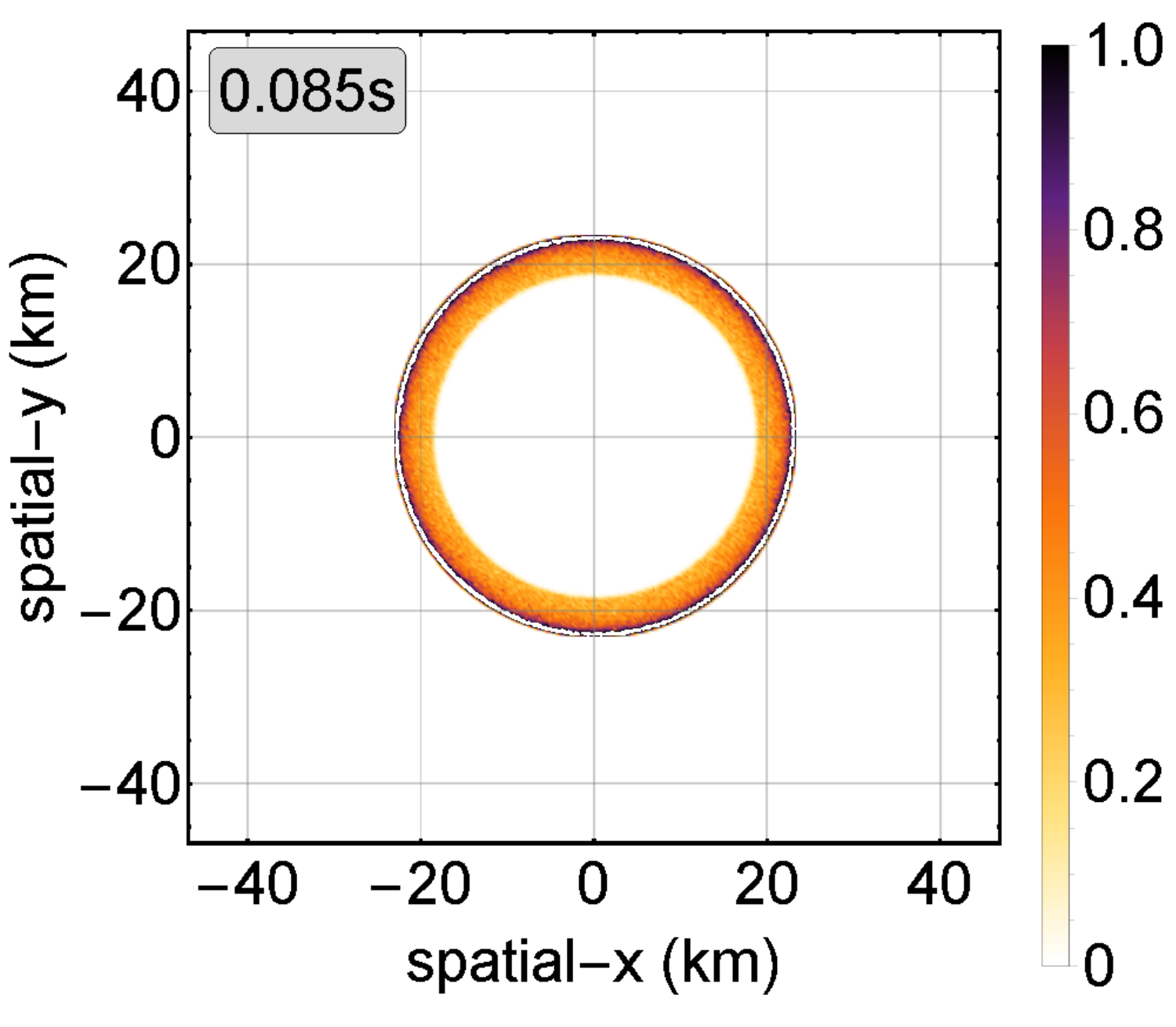} & \includegraphics[scale=0.17]{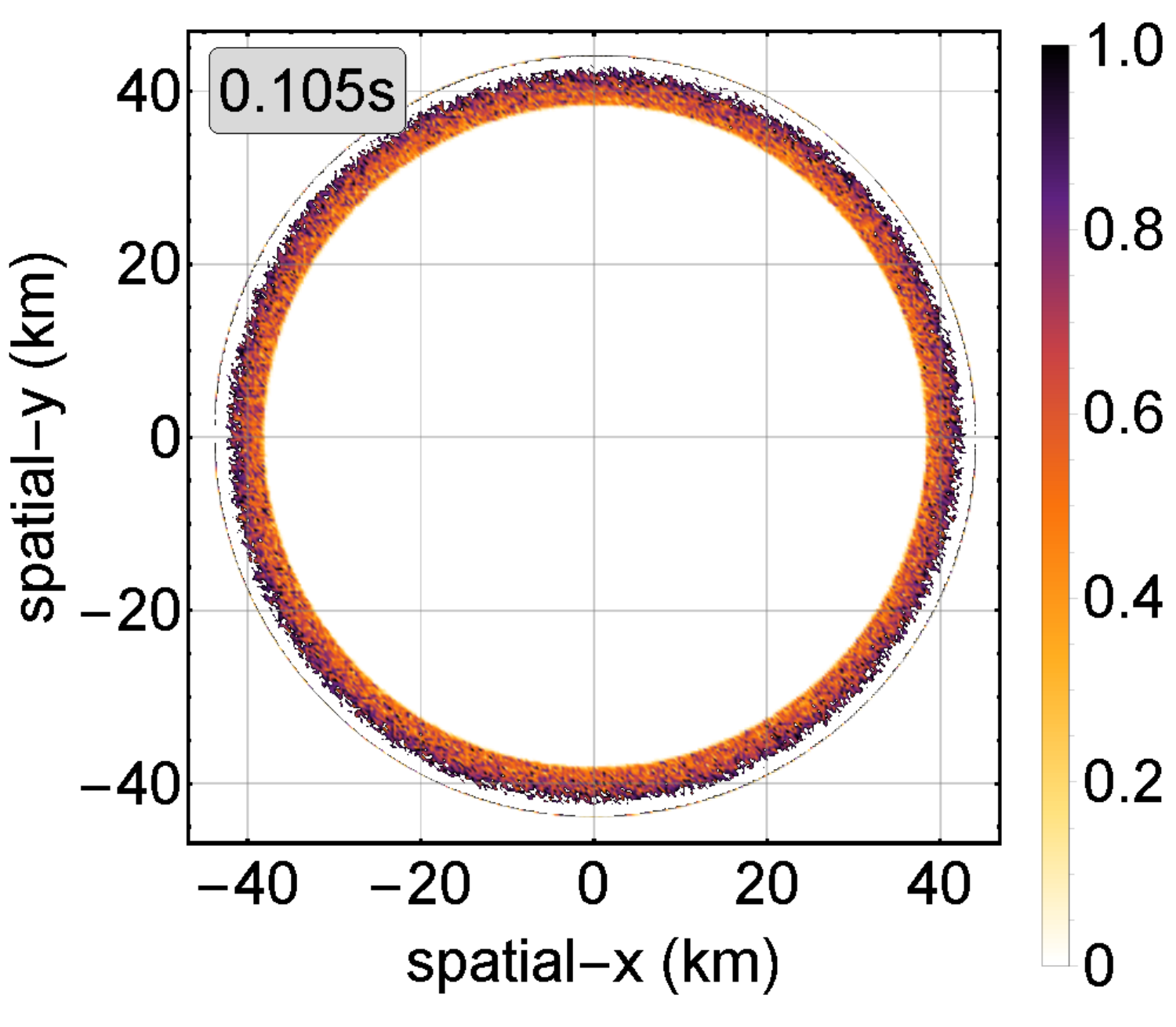} & \includegraphics[scale=0.17]{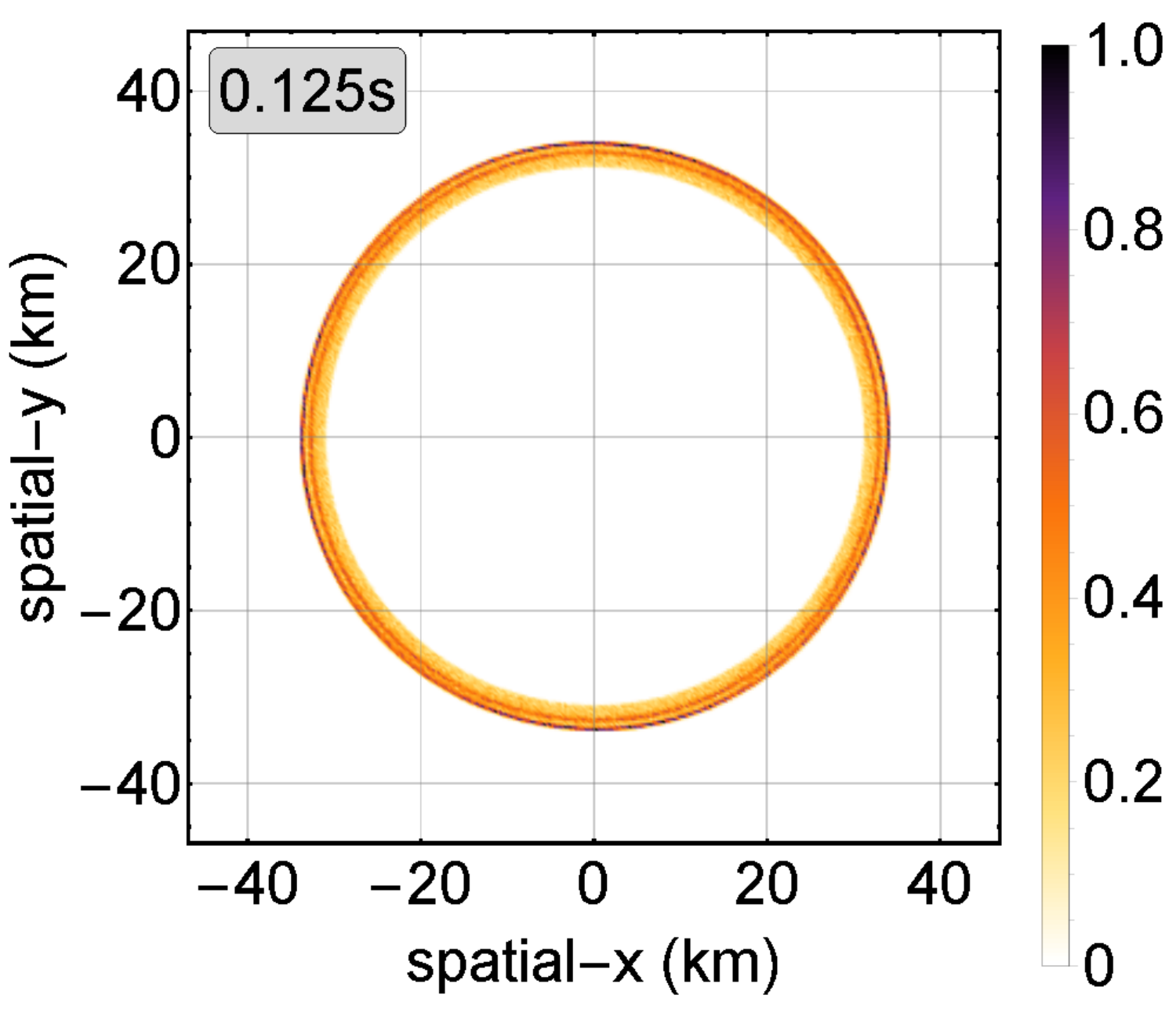} & \includegraphics[scale=0.17]{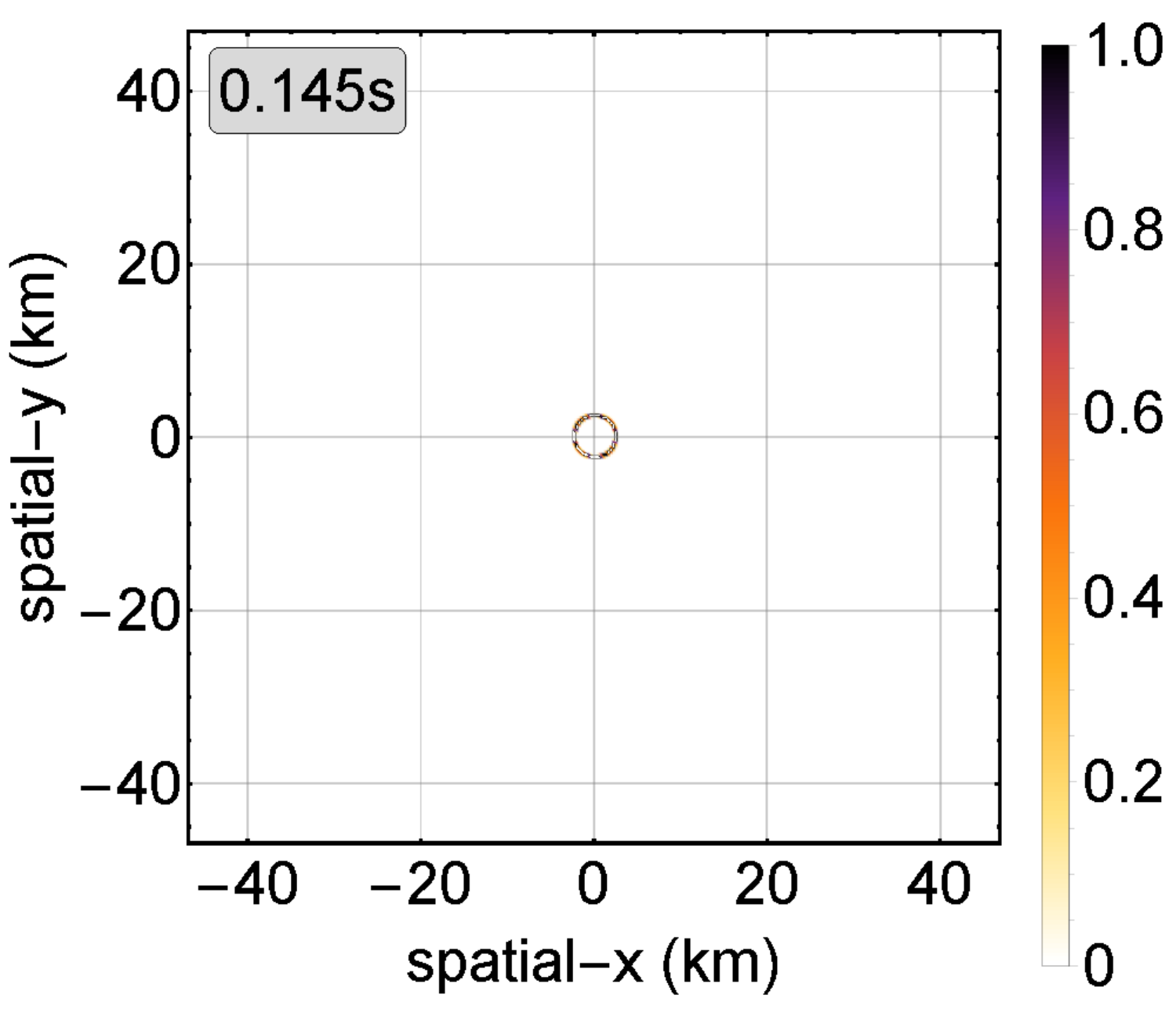}\tabularnewline
\end{tabular}
\par\end{centering}
\caption{Evolution of the heavy ions at various snapshots with $\Gamma\simeq1.09$.
The figures are viewed in the direction of the ambient magnetic field.
The total mass of the initial heavy ions is taken to be $0.02\ \text{kg}$.
Other parameters are the same as in Fig. \ref{fig:Evolution-of-the}.
We can see the breathing patterns of the ions in this case. \label{fig:Evolution-of-the-1}}

\end{figure}

\begin{figure}
\begin{centering}
\begin{tabular}{cccc}
\includegraphics[scale=0.17]{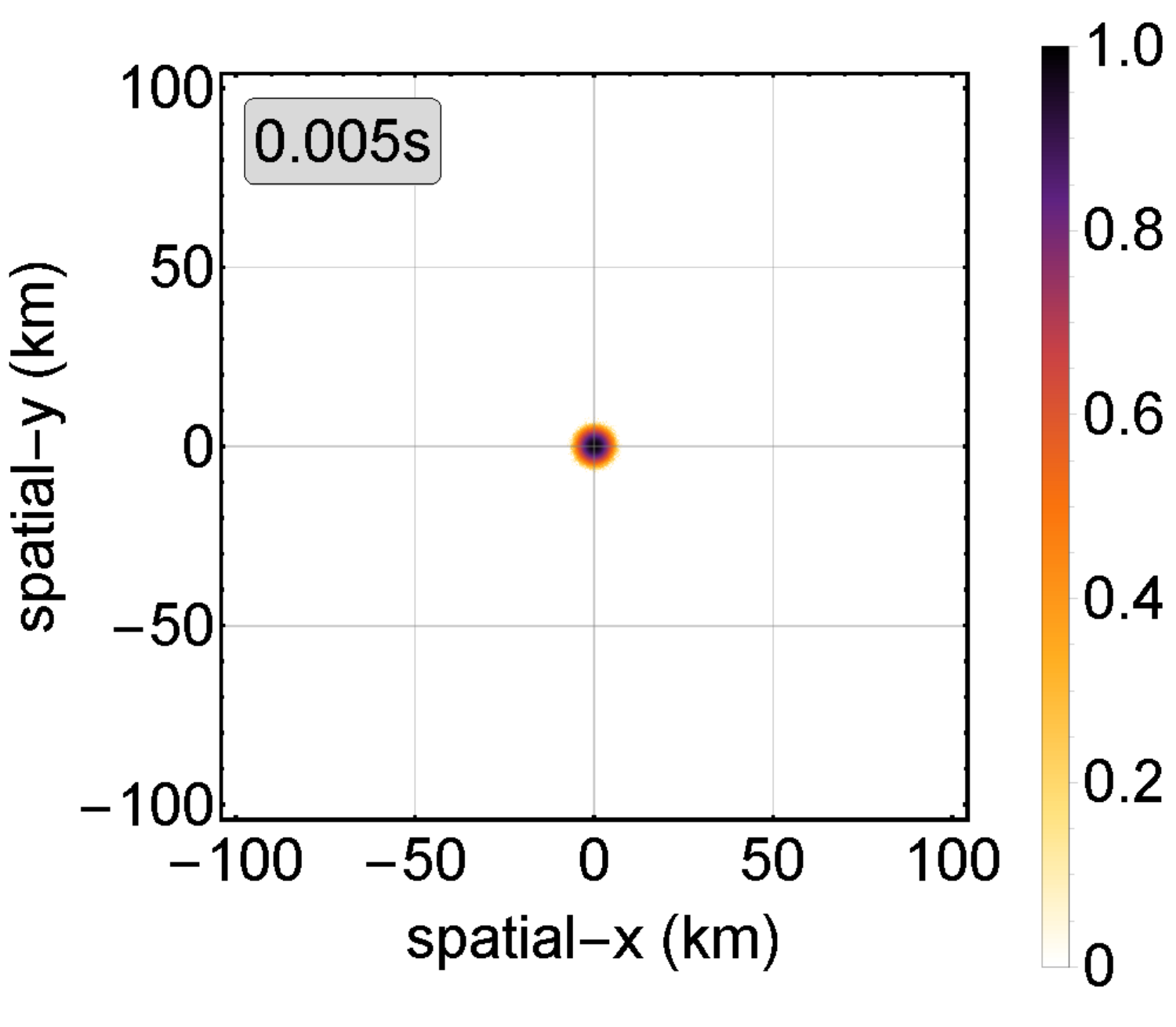} & \includegraphics[scale=0.17]{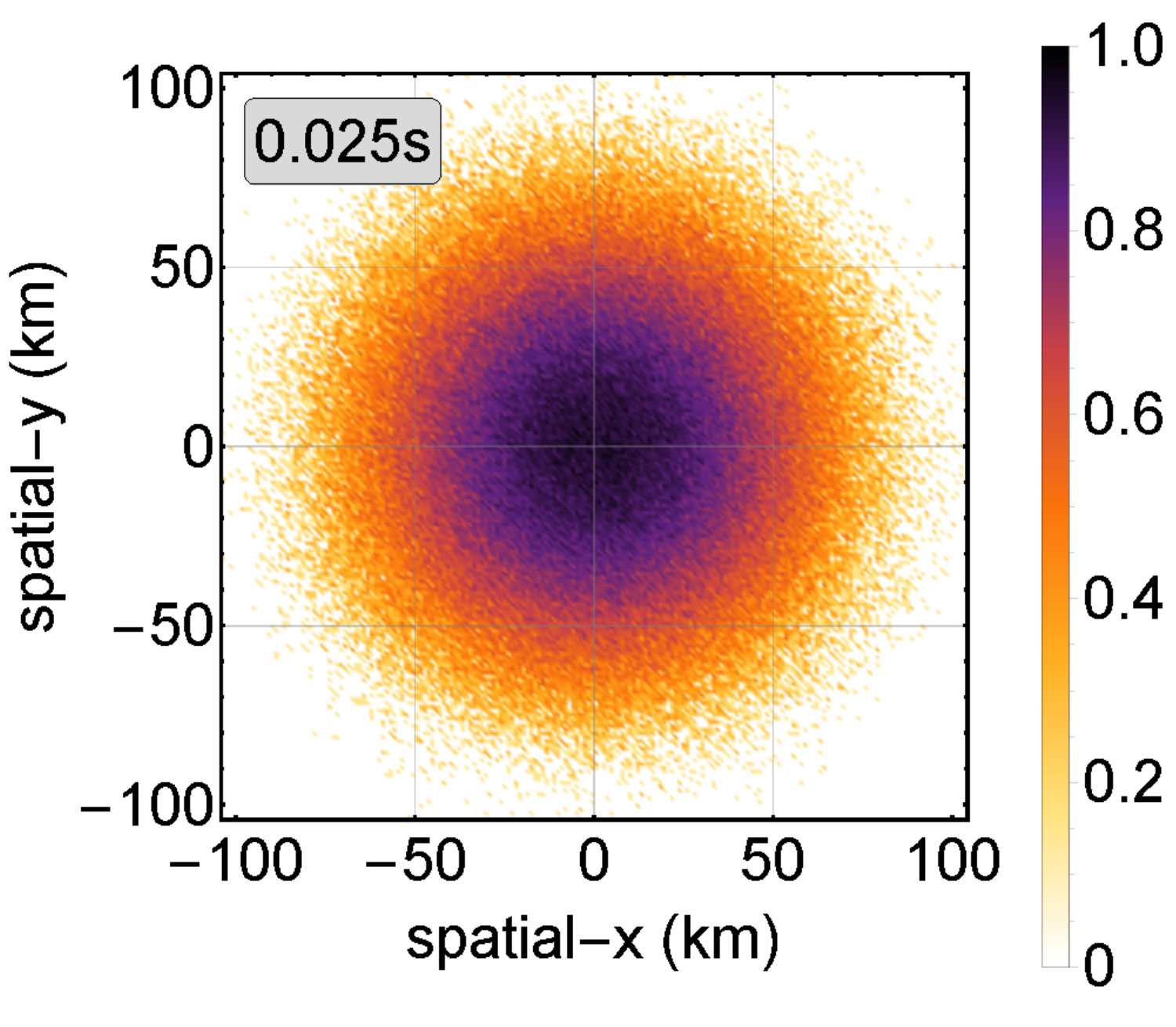} & \includegraphics[scale=0.17]{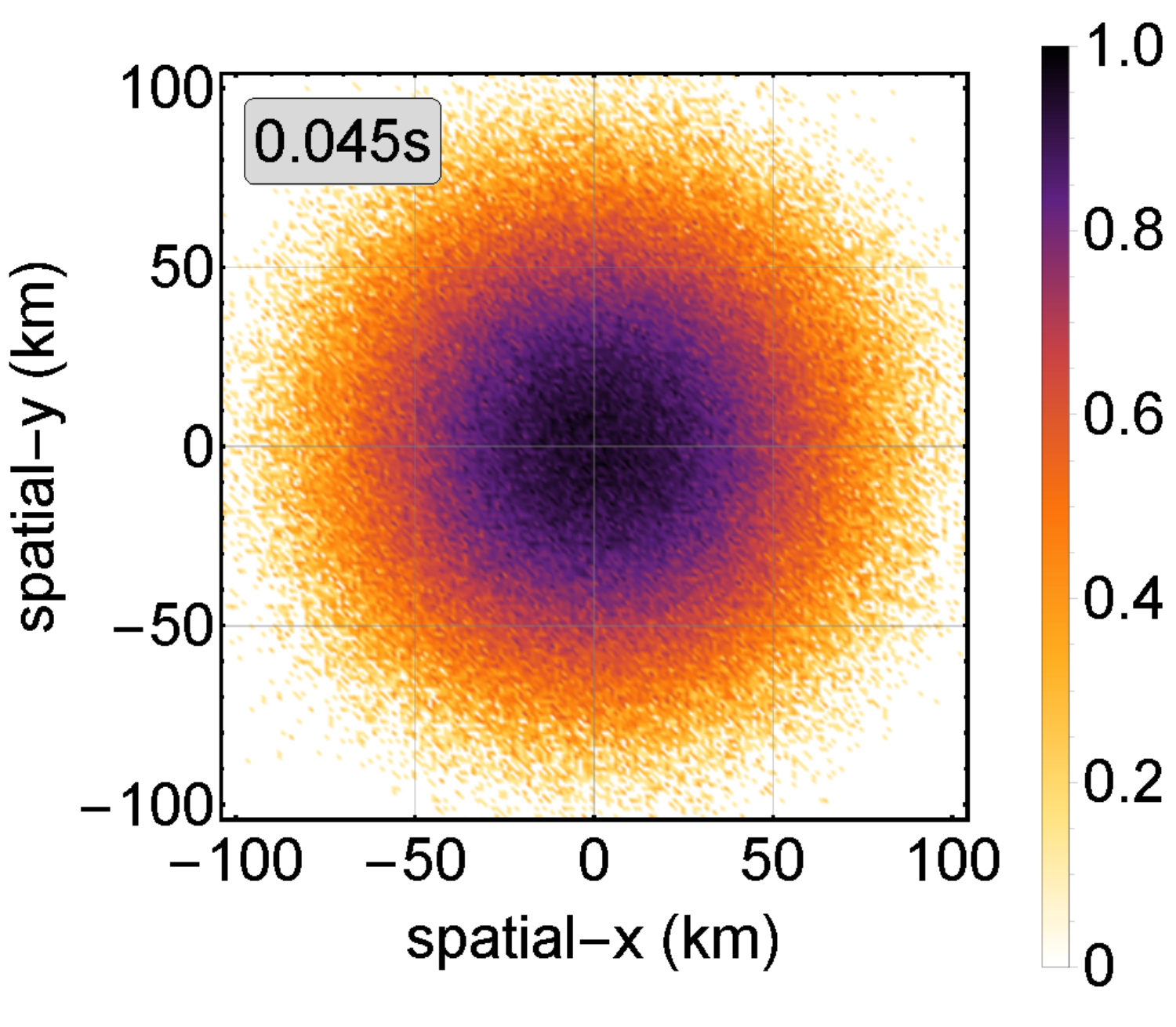} & \includegraphics[scale=0.17]{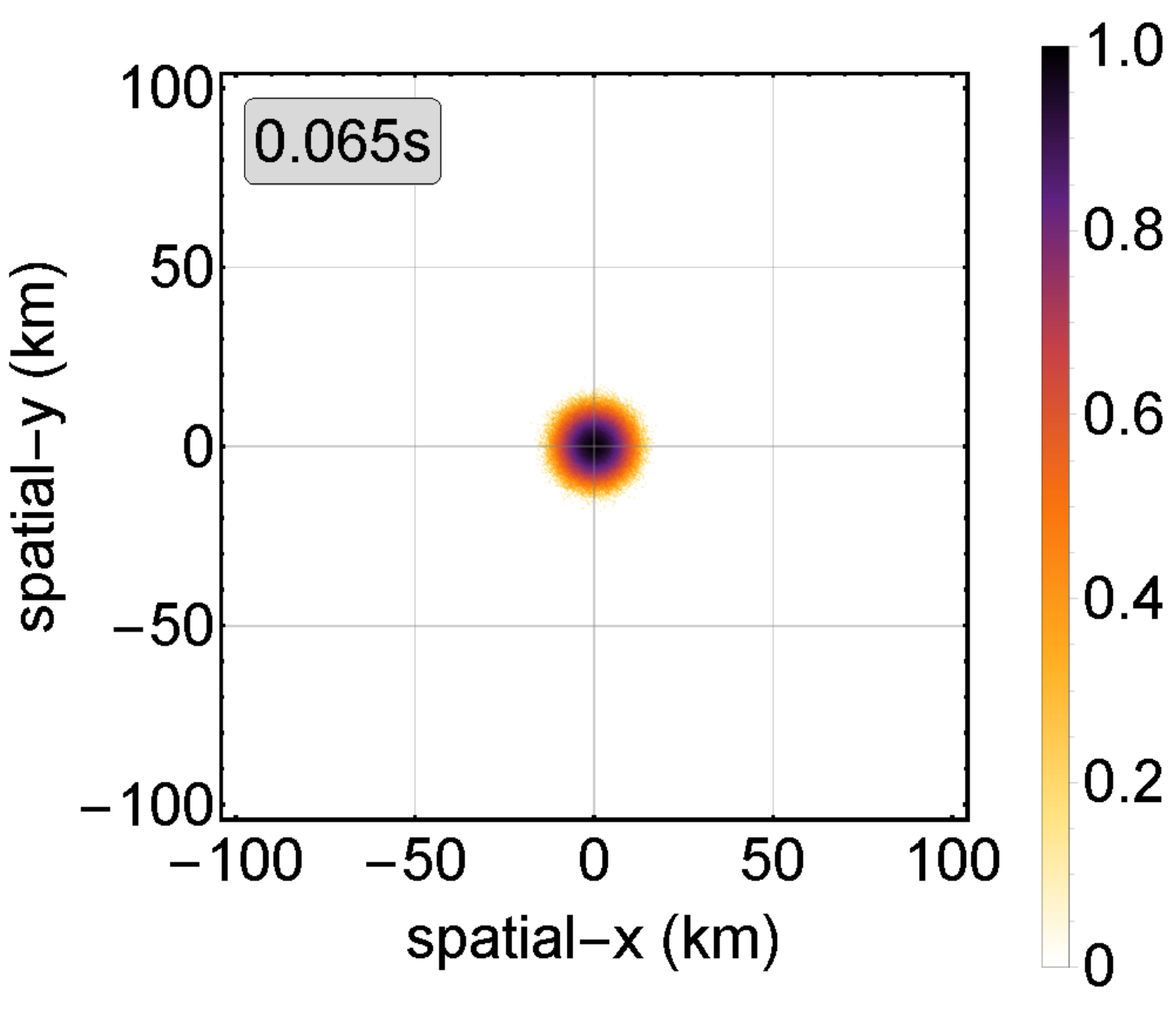}\tabularnewline
\includegraphics[scale=0.17]{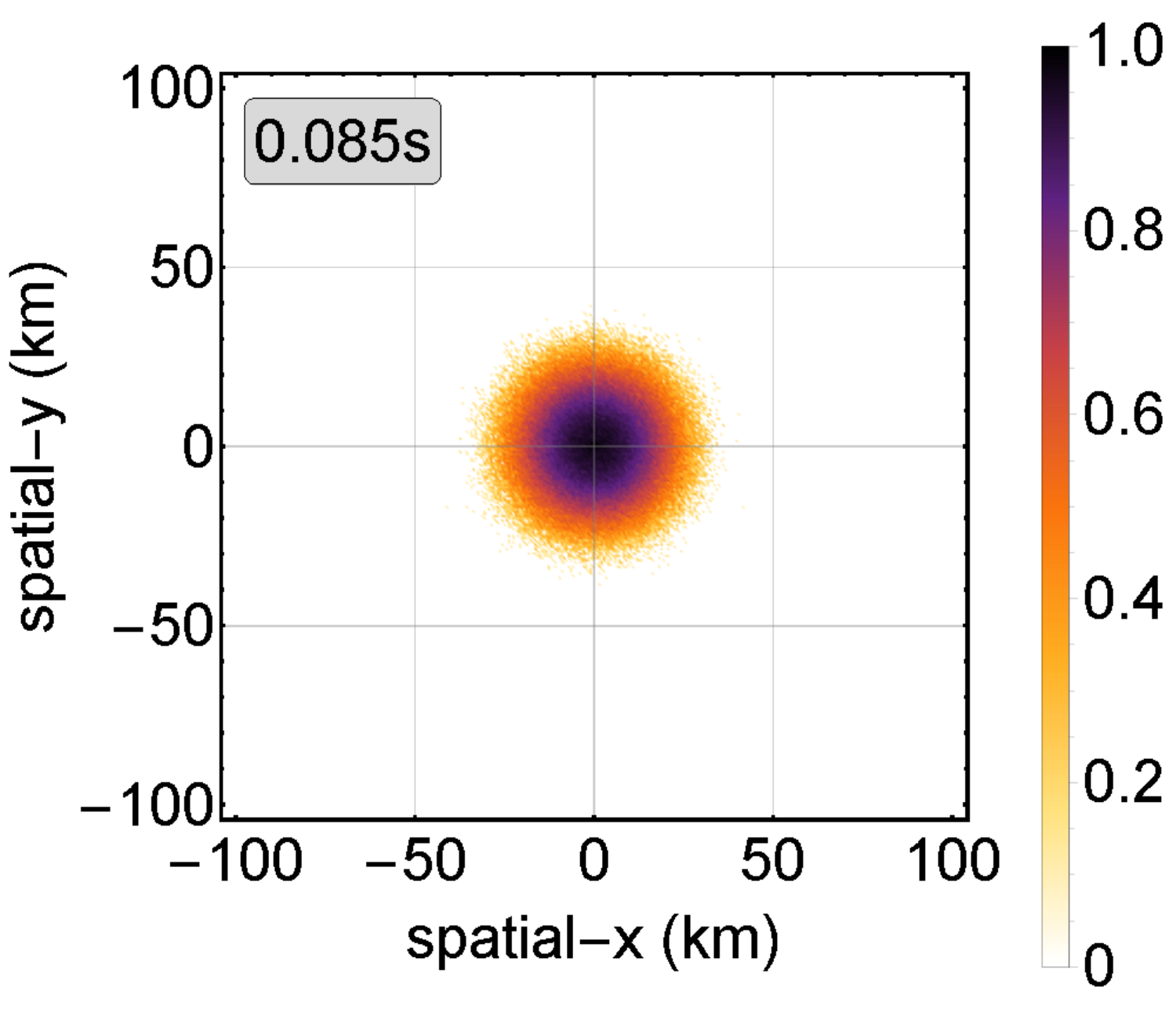} & \includegraphics[scale=0.17]{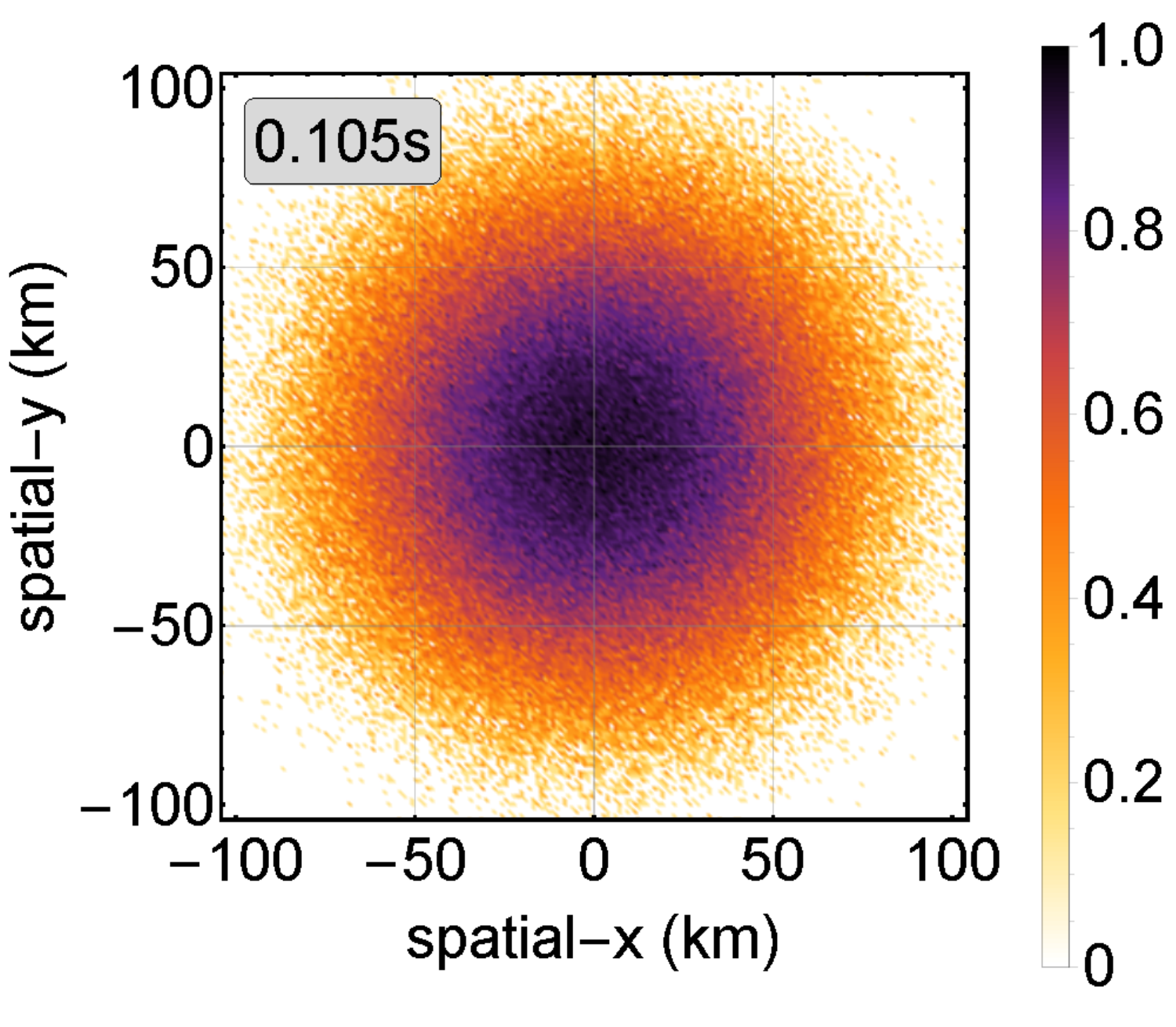} & \includegraphics[scale=0.17]{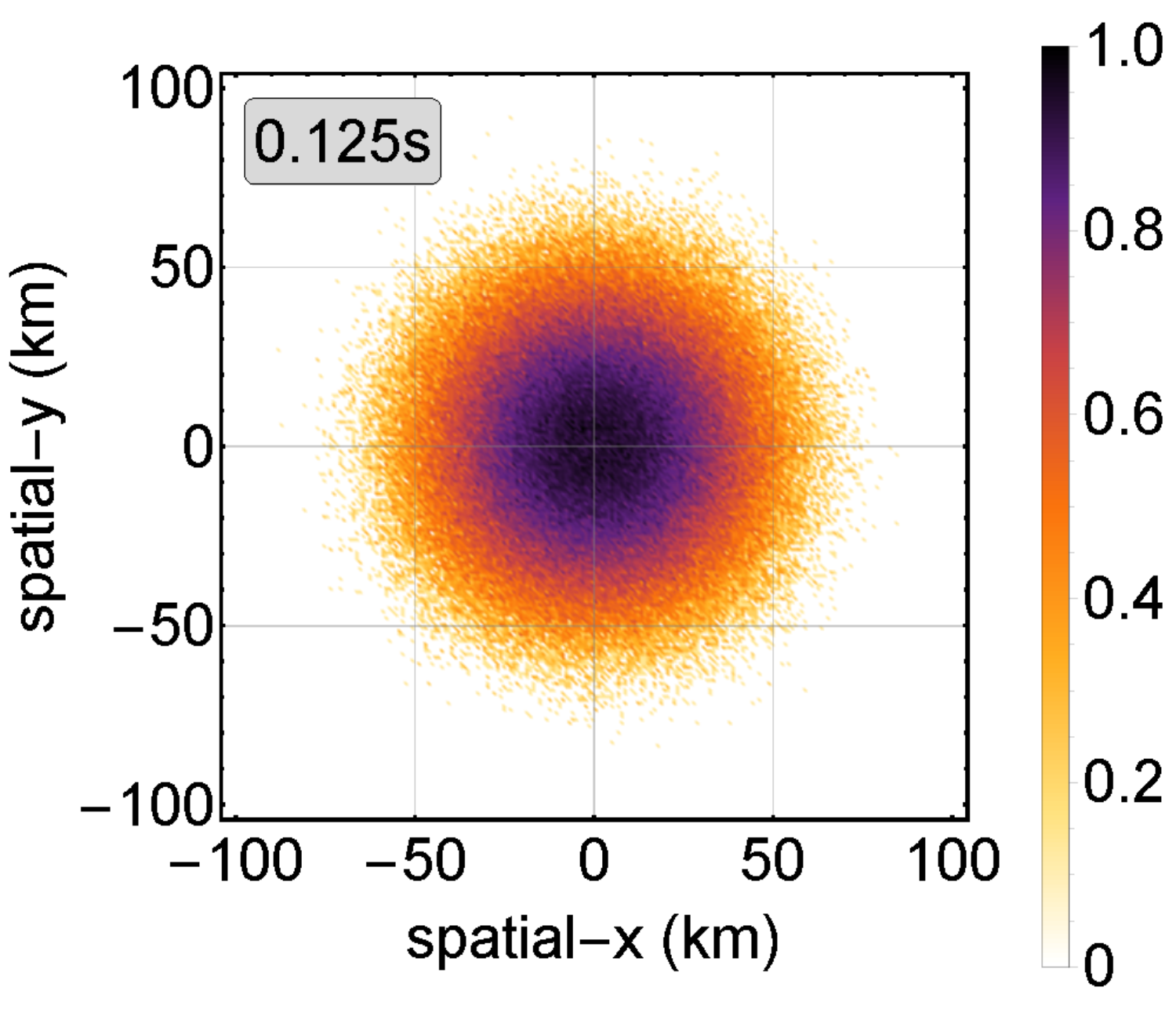} & \includegraphics[scale=0.17]{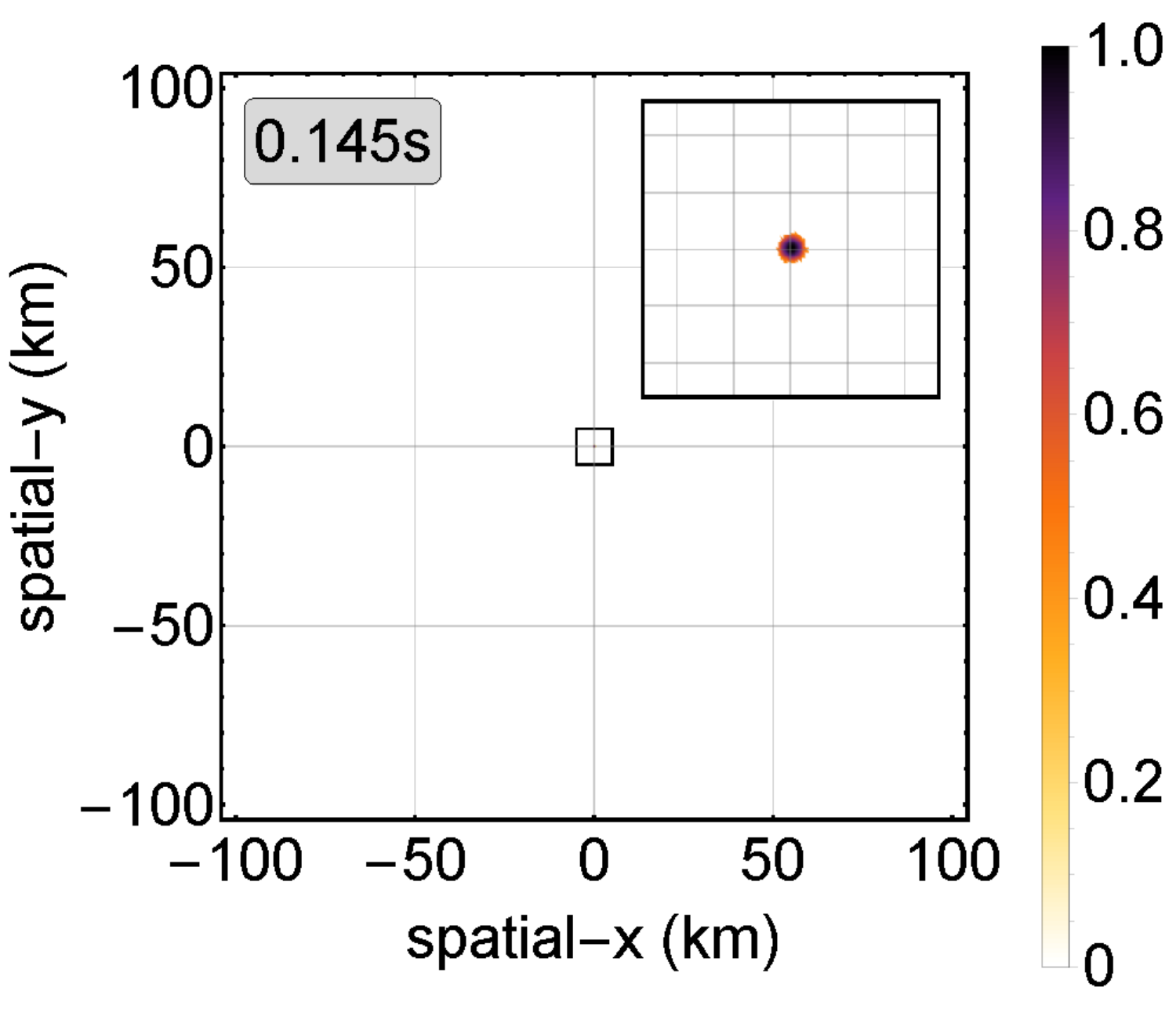}\tabularnewline
\end{tabular}
\par\end{centering}
\caption{Evolution of the heavy ions at various snapshots with $\Gamma\simeq1.09$.
The figures are viewed in the direction of the ambient magnetic field.
The total mass of the initial ions is taken to be $0.02\ \text{kg}$.
Initial velocity of the ions is sampled from a Maxwellian distribution
with average velocity $\overline{v}_{i,\text{Fe}}=2\times10^{6}\ \text{m/s}$.
Other parameters are set the same as in Fig. \ref{fig:Evolution-of-the}.\label{fig:Evolution-of-the-2}}

\end{figure}

The two typical patterns of the heavy ions can be determined by their
initial total mass. The difference in the total mass will lead to
different $\Gamma$ values defined in Sec. \ref{sec:The-simulation-model}.
We will demonstrate the simulation results of two values $\Gamma=1.09$
(Fig. \ref{fig:Evolution-of-the-1} and \ref{fig:Evolution-of-the-2})
and $\Gamma=0.03$ (Fig. \ref{fig:Evolution-of-the}) which correspond
to the laser experiments and the near space plasma. 

Fig. \ref{fig:Evolution-of-the-1} gives the evolution of the heavy
ions when $\Gamma\simeq1.09$. In such a condition like in the laser
experiments, we see a very different pattern of the ion-motion compared
with that in Fig. \ref{fig:Evolution-of-the}. The heavy ions initially
expand outward and begin to form a thin shell. Quite unexpectedly,
these ions then stop expanding and shrink back into the burst point.
This breathing pattern continues until most of the ions lose their
energy to the background plasma and finally stop moving. 

\begin{figure}
\centering{}%
\begin{tabular}{ccc}
\includegraphics[scale=0.24]{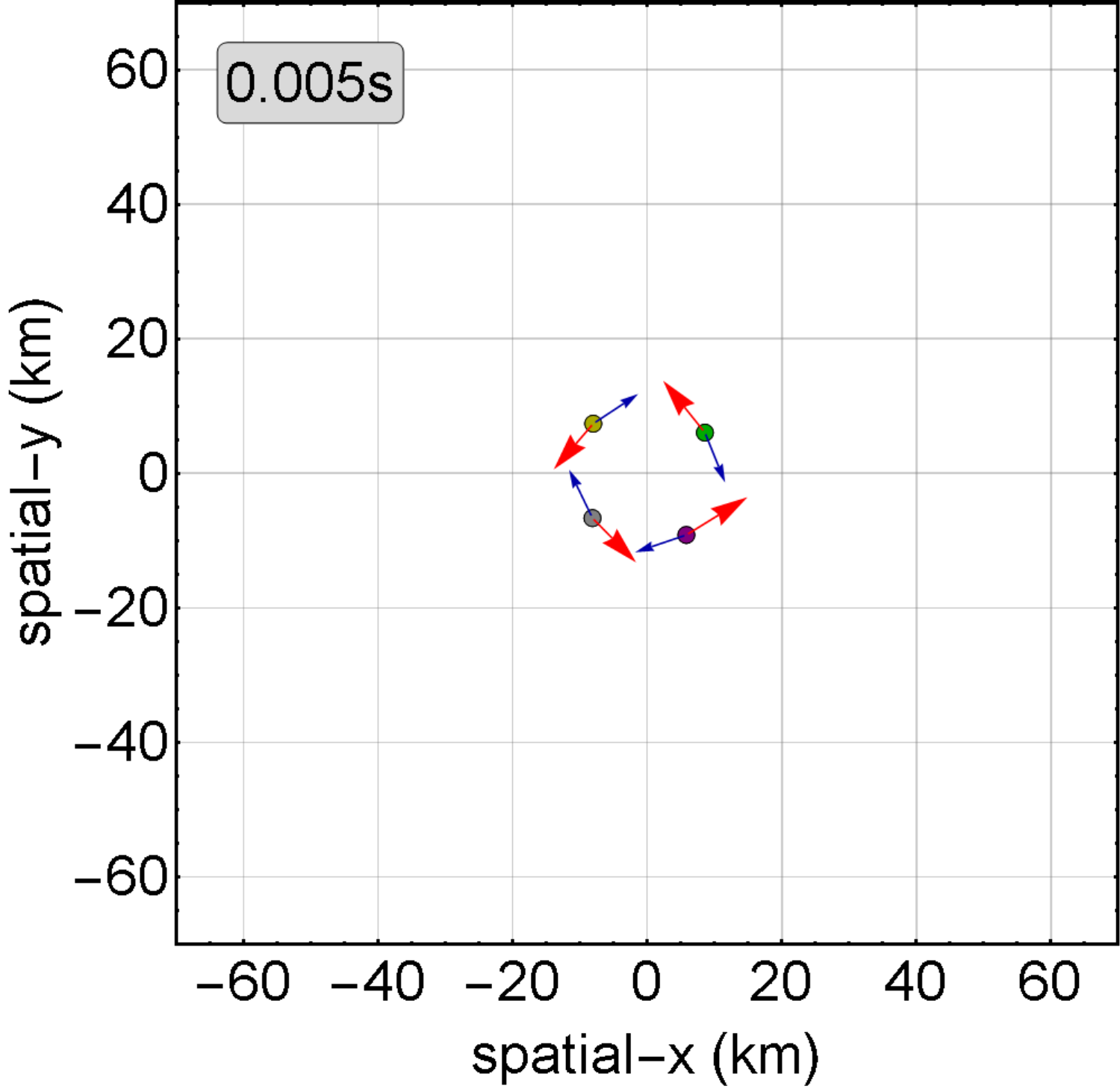} & \includegraphics[scale=0.24]{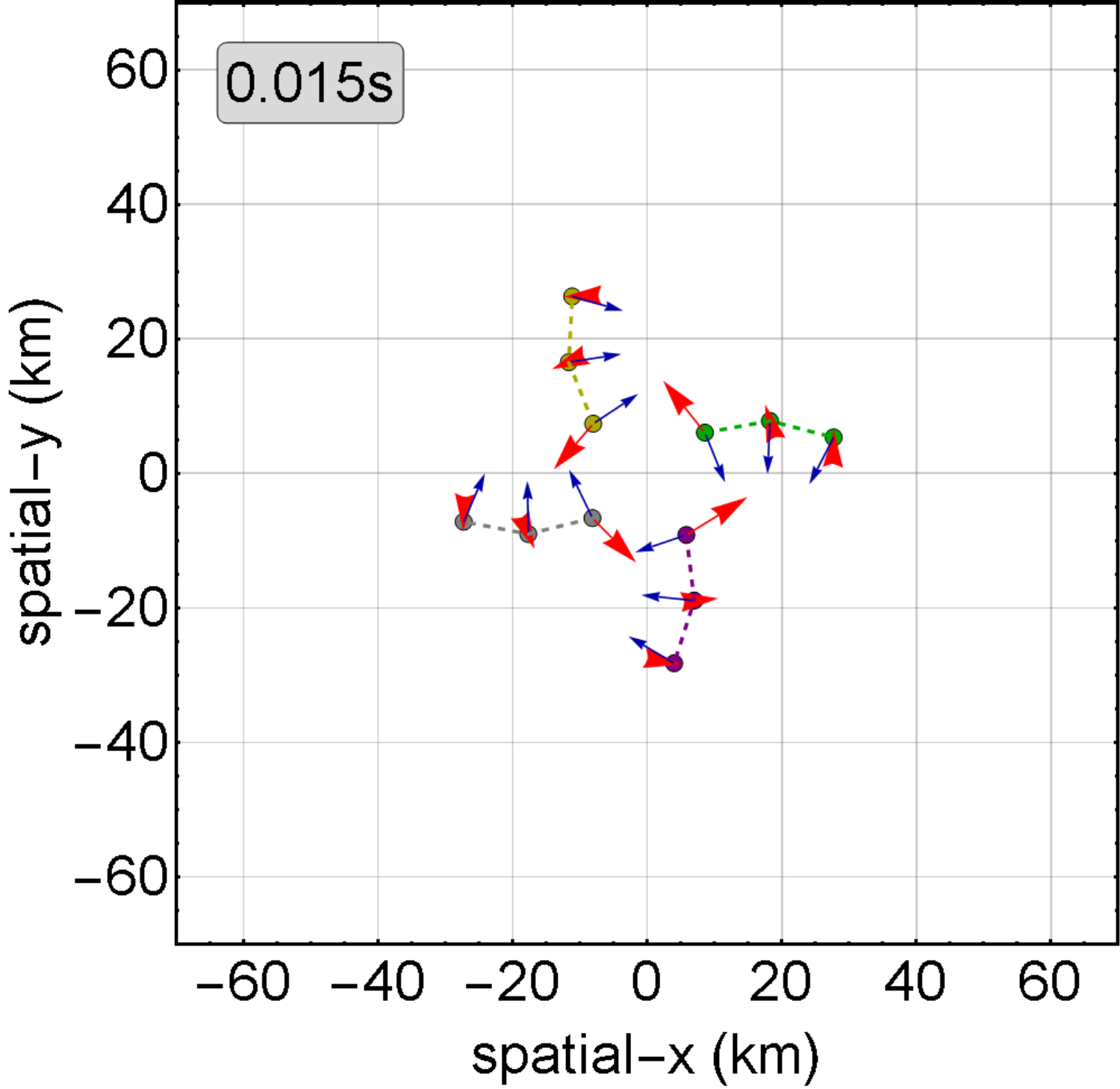} & \includegraphics[scale=0.24]{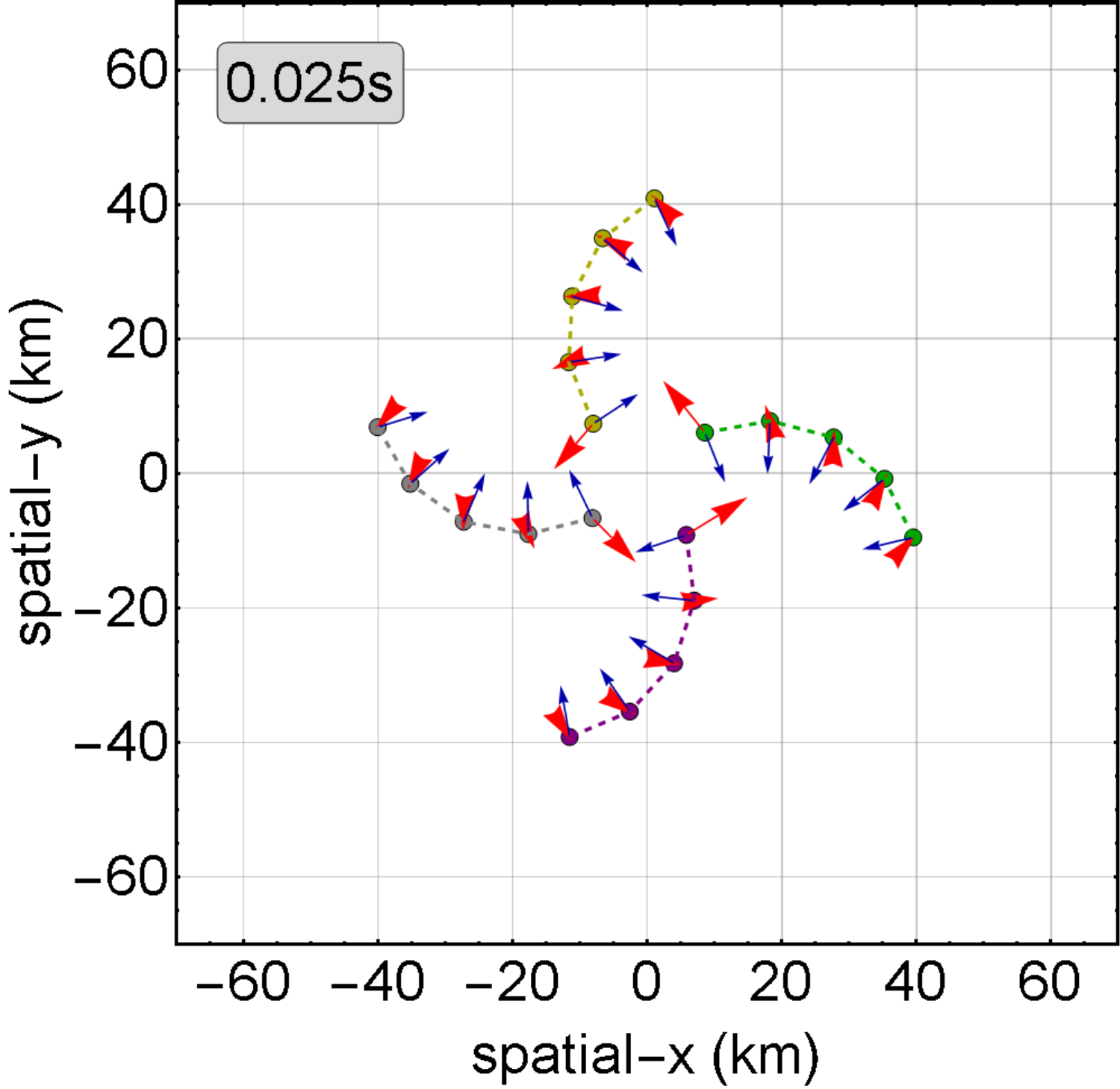}\tabularnewline
\includegraphics[scale=0.24]{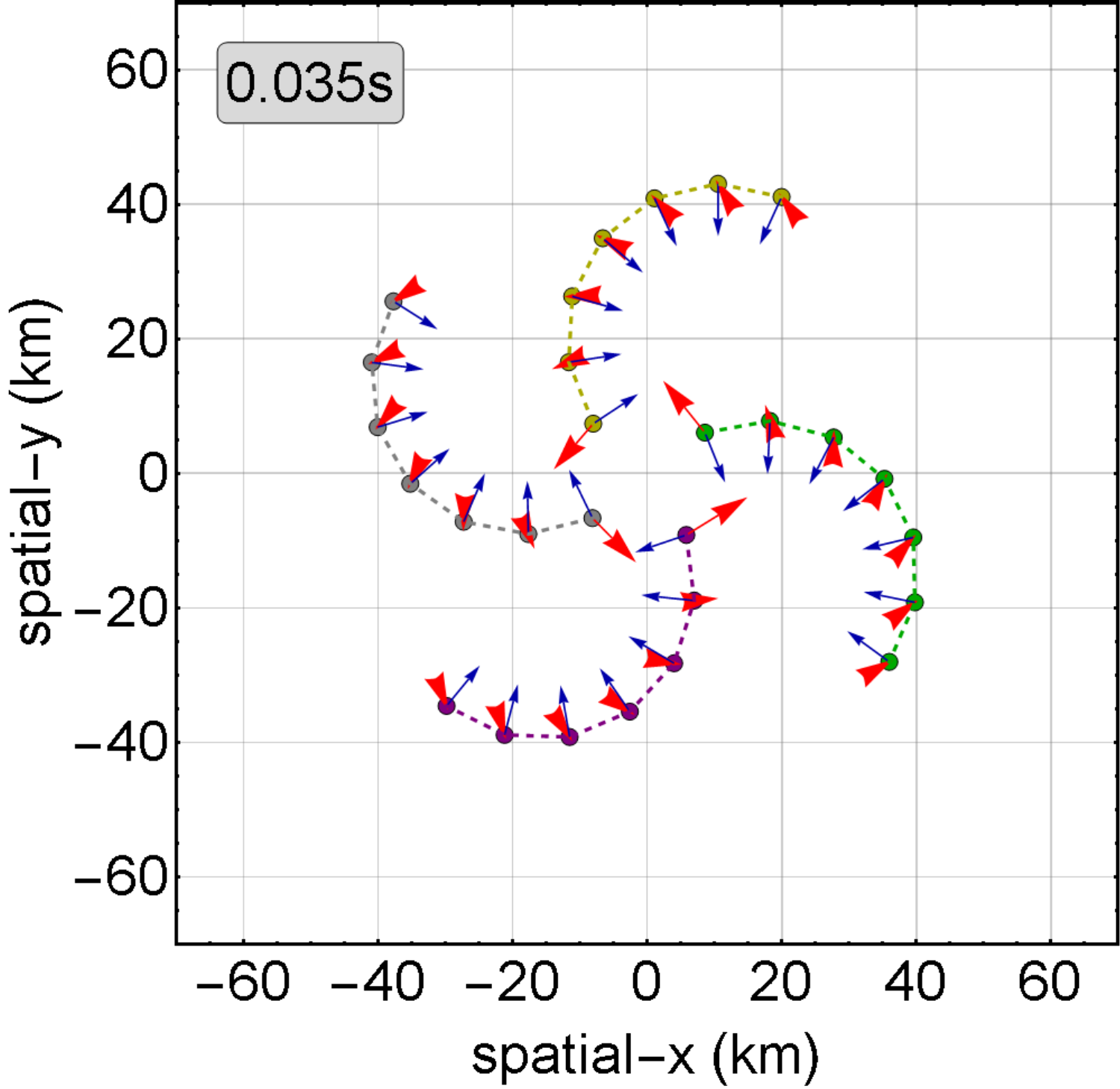} & \includegraphics[scale=0.24]{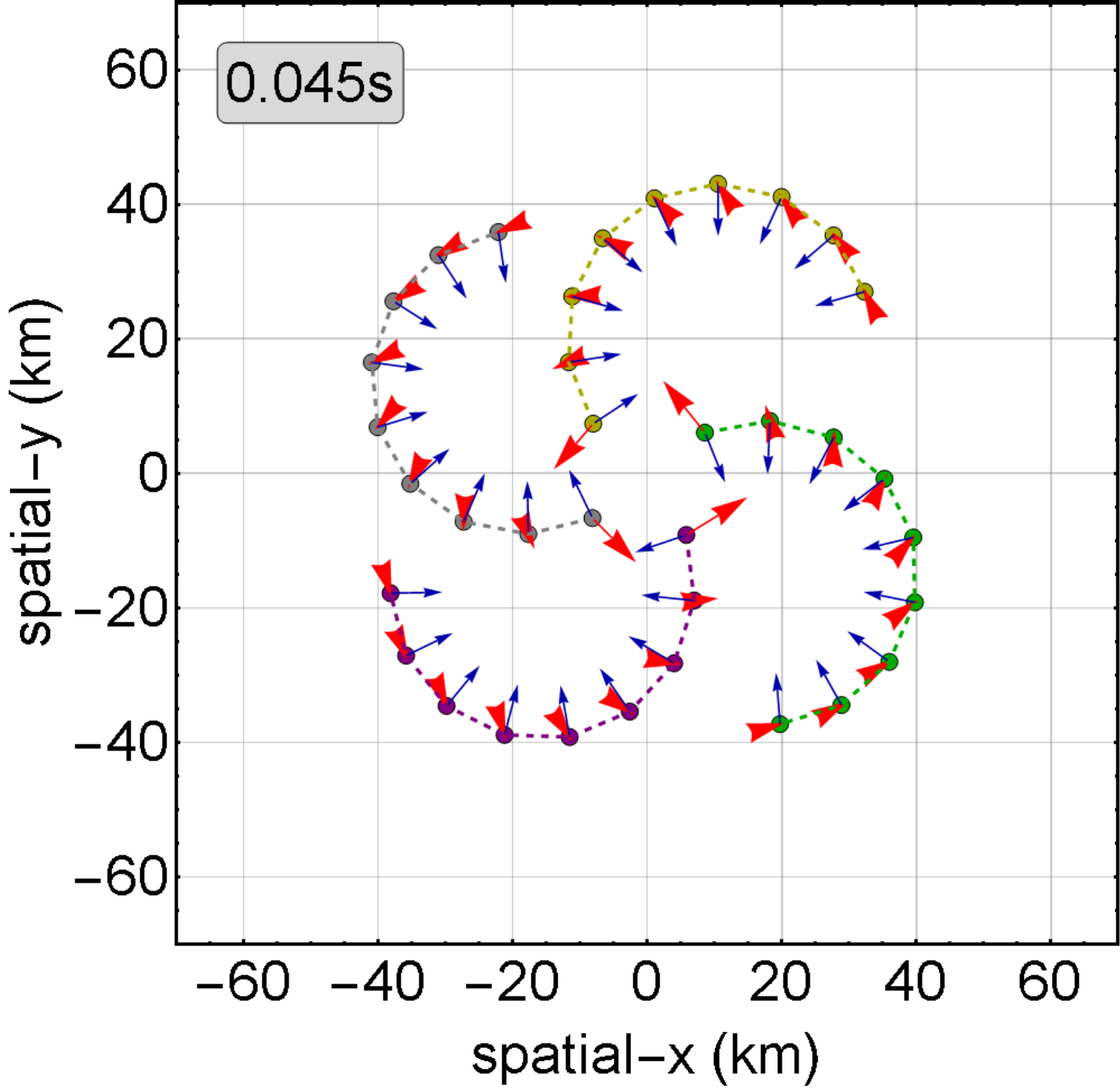} & \includegraphics[scale=0.24]{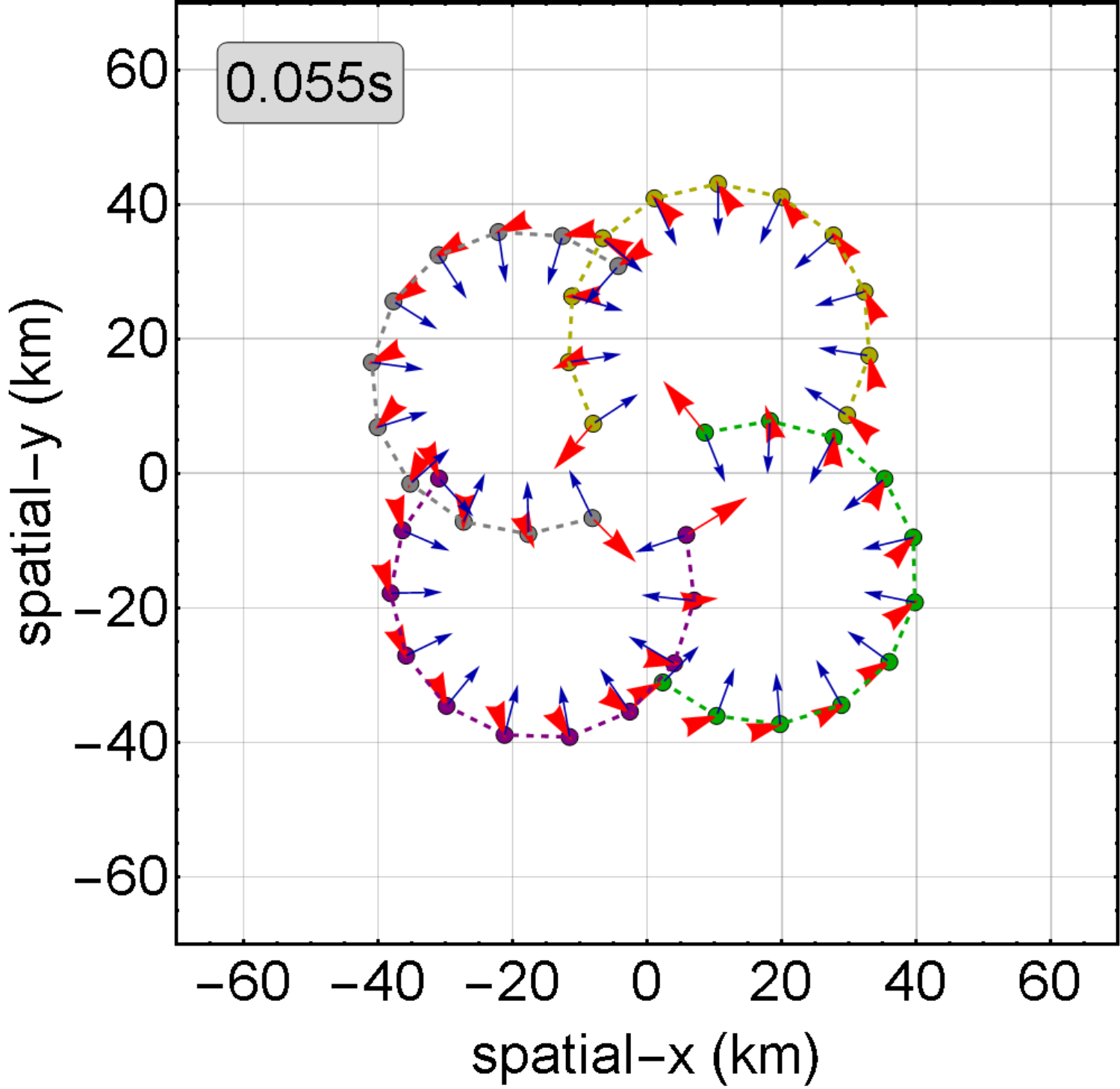}\tabularnewline
\end{tabular}\caption{Motion of the selected debris particles at various snapshots when
$\Gamma\simeq1.09$. The parameters are set the same as in Fig. \ref{fig:Evolution-of-the-1}.
The red and blue arrows in each graph indicate the directions of the
electric and magnetic (Lorentz) force received by the particles, respectively.
The trajectories of these particles are shown in different colors.\label{fig:Motion-of-the}}
\end{figure}

To see exactly what happens to these ions, we have picked four ion
particles and tracked their motions at each snapshot. As is depicted
in Fig. \ref{fig:Motion-of-the}, the Lorentz force is always perpendicular
to the motion of the particles. However, the electric force, which
is initially perpendicular to the trajectories of the particles, gradually
tilts parallel to the velocity of the heavy ions at the rim of the
thin shell. The parallel electric force is the main contributor that
decreases the energy of the heavy ions and meanwhile generates a circular
electric field that propagates outward.

One may suspect that this breathing pattern is caused by the initial
set-up of constant radial velocity since the collective motion of
these ions is sensitive to their initial velocity. In fact, the breathing
pattern is mainly determined by the initial mass of the heavy ions
and is not so sensitive to the initial distribution of their velocities.
To perceive this argument, we have demonstrated a case in which the
ions initial velocities are sampled from a Maxwellian distribution,
as in Fig. \ref{fig:Evolution-of-the-2}. We see clearly that the
breathing pattern still dominates the evolution of the heavy ions
except for an absence of a thin shell. 

\begin{figure}
\begin{centering}
\begin{tabular}{ccc}
\includegraphics[scale=0.24]{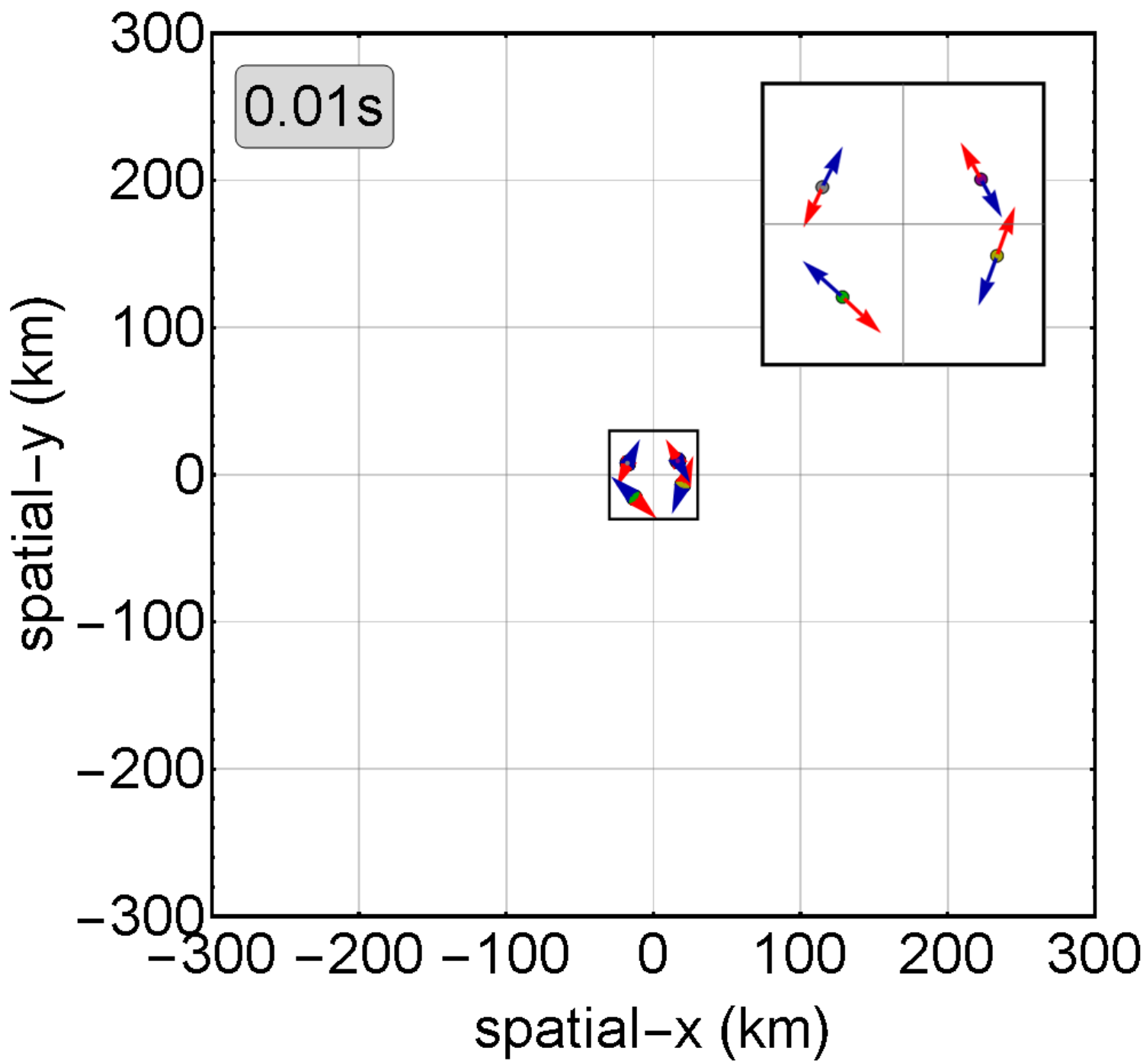} & \includegraphics[scale=0.24]{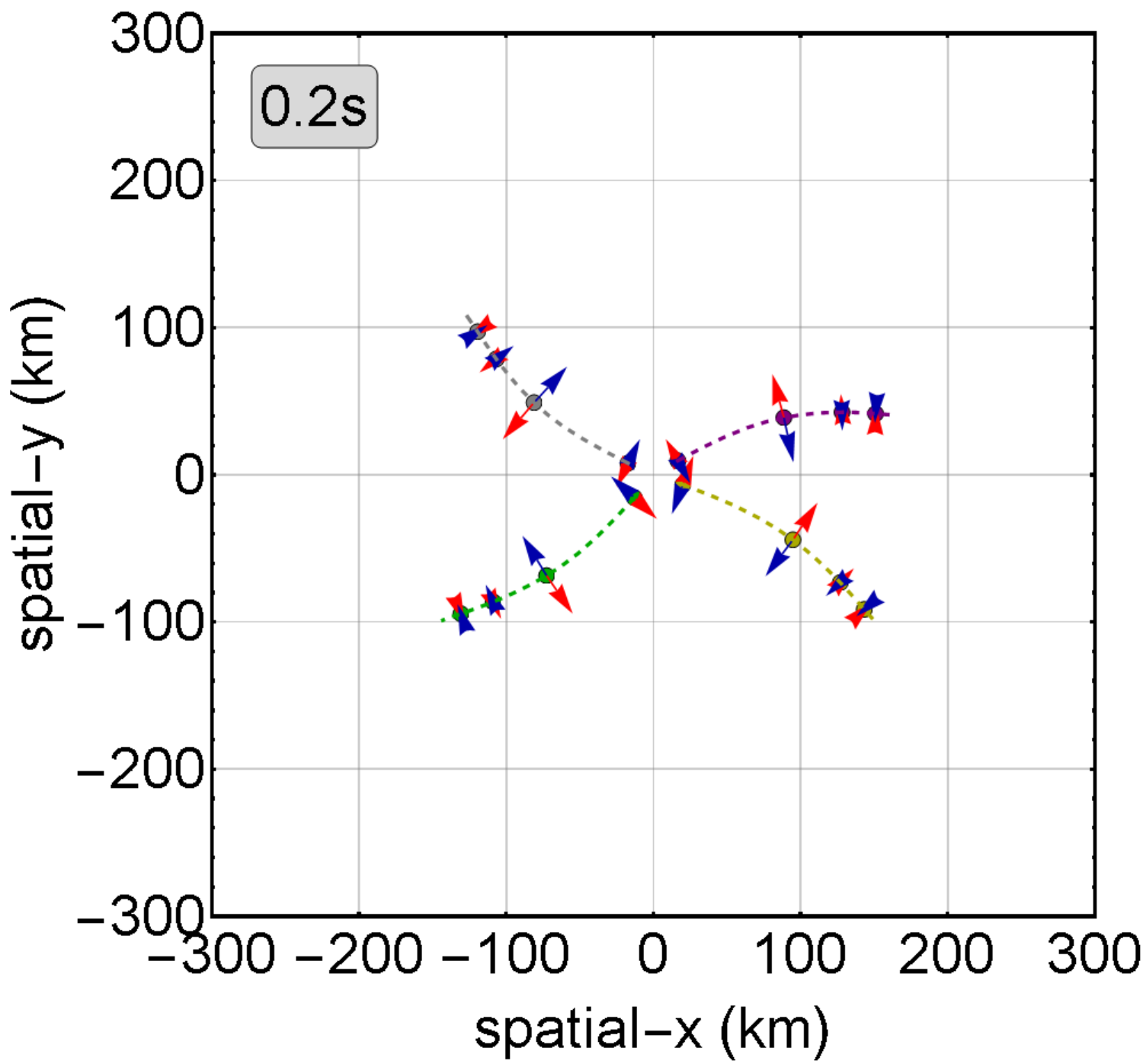} & \includegraphics[scale=0.24]{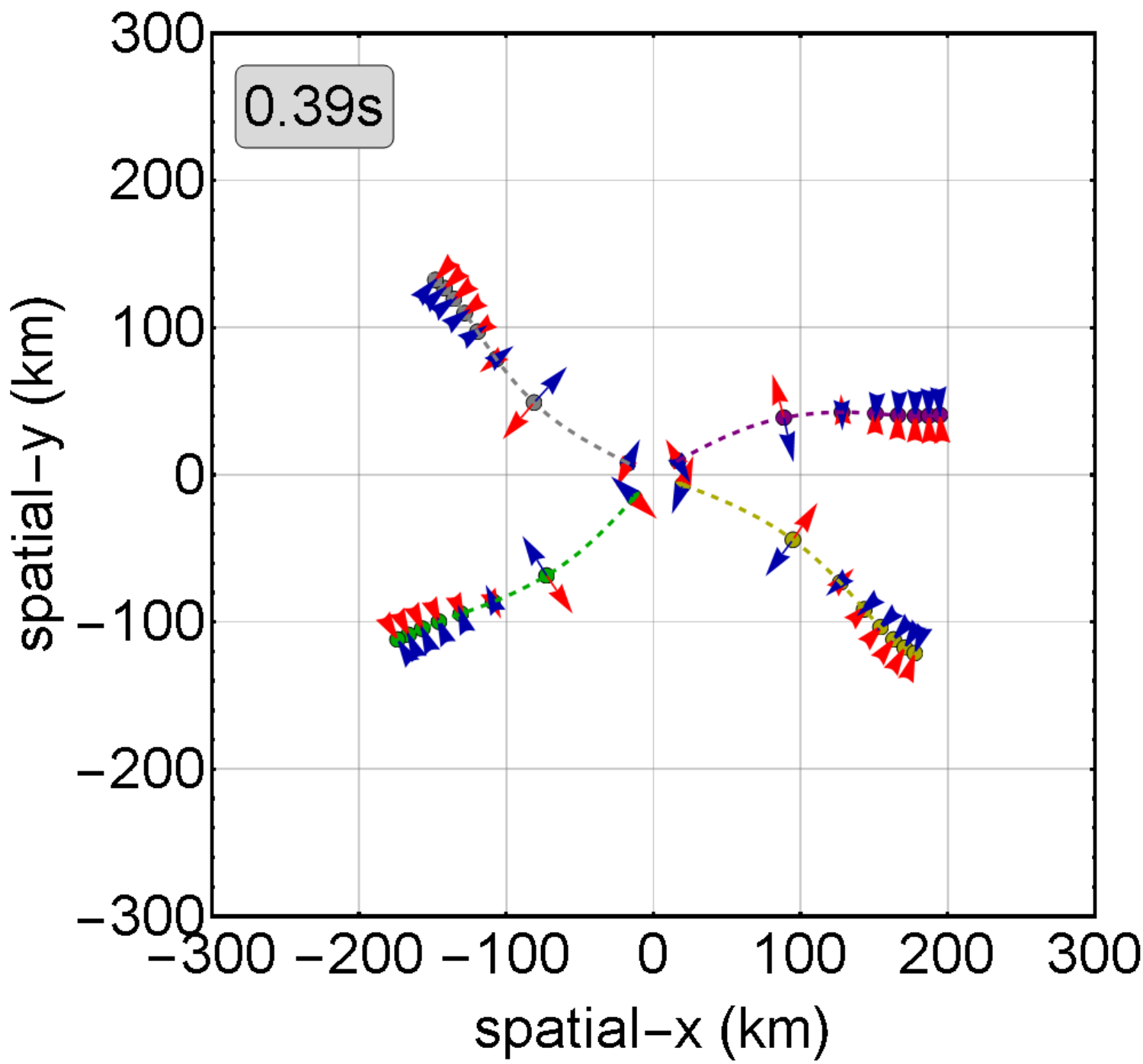}\tabularnewline
\includegraphics[scale=0.24]{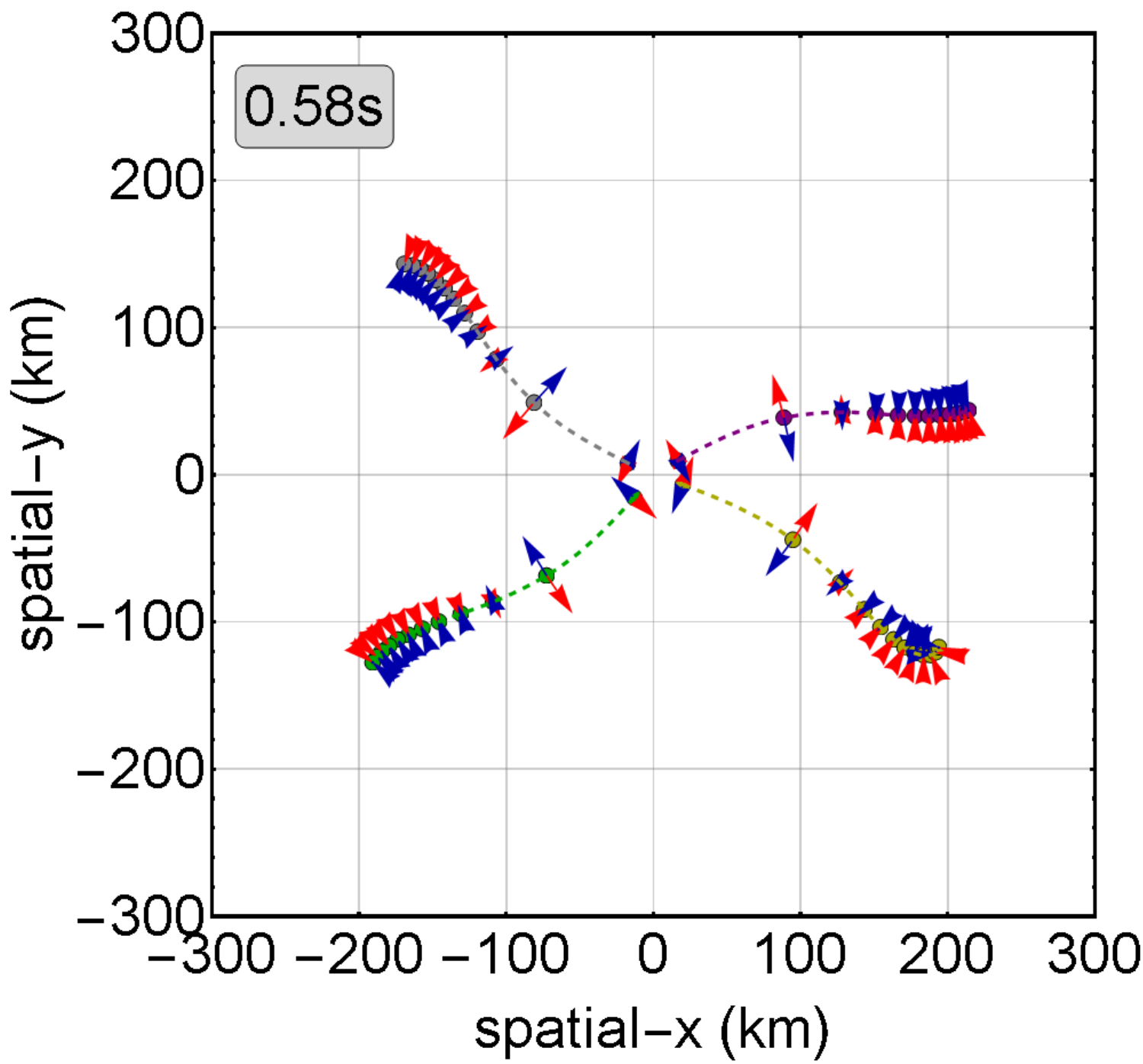} & \includegraphics[scale=0.24]{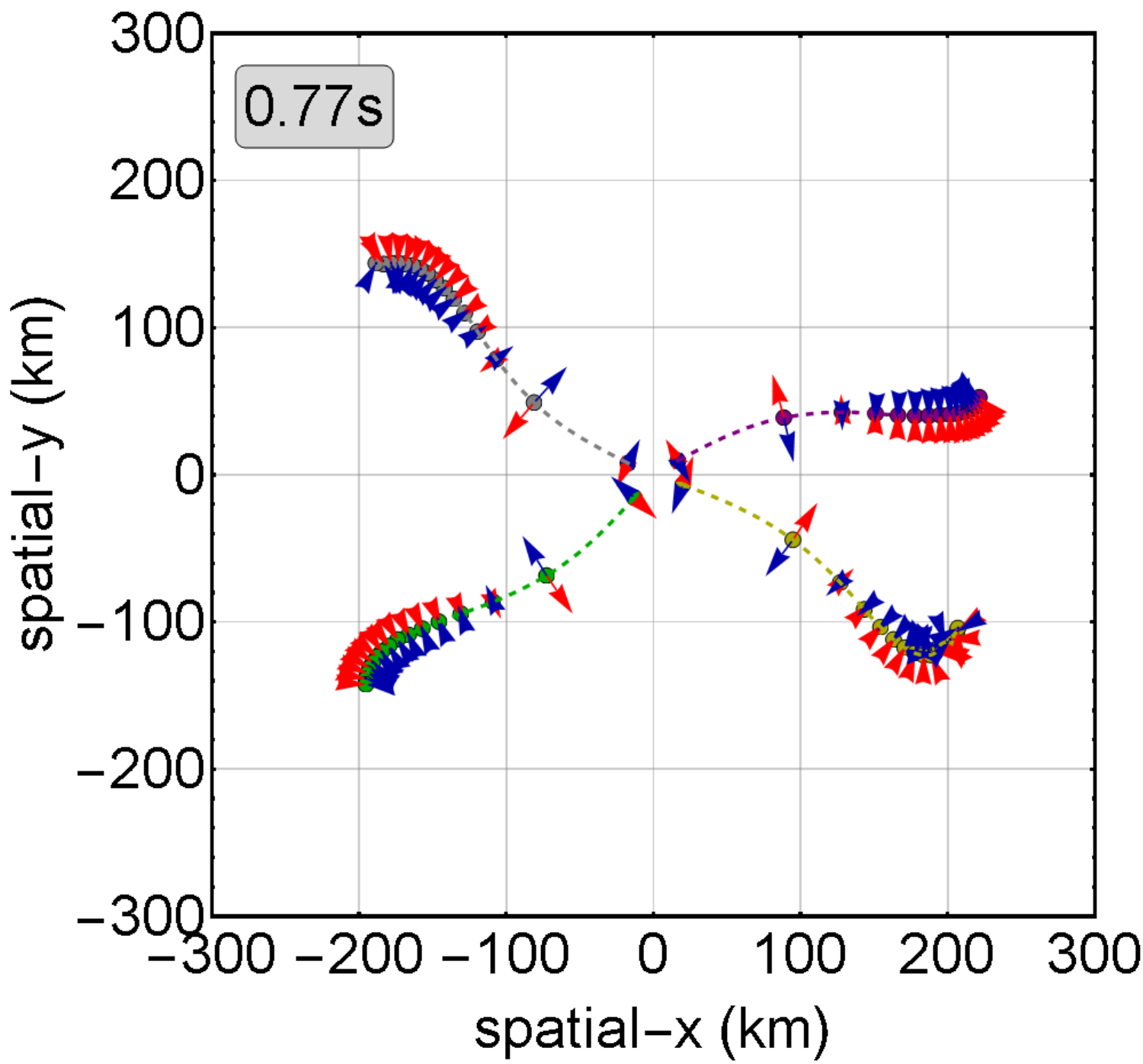} & \includegraphics[scale=0.24]{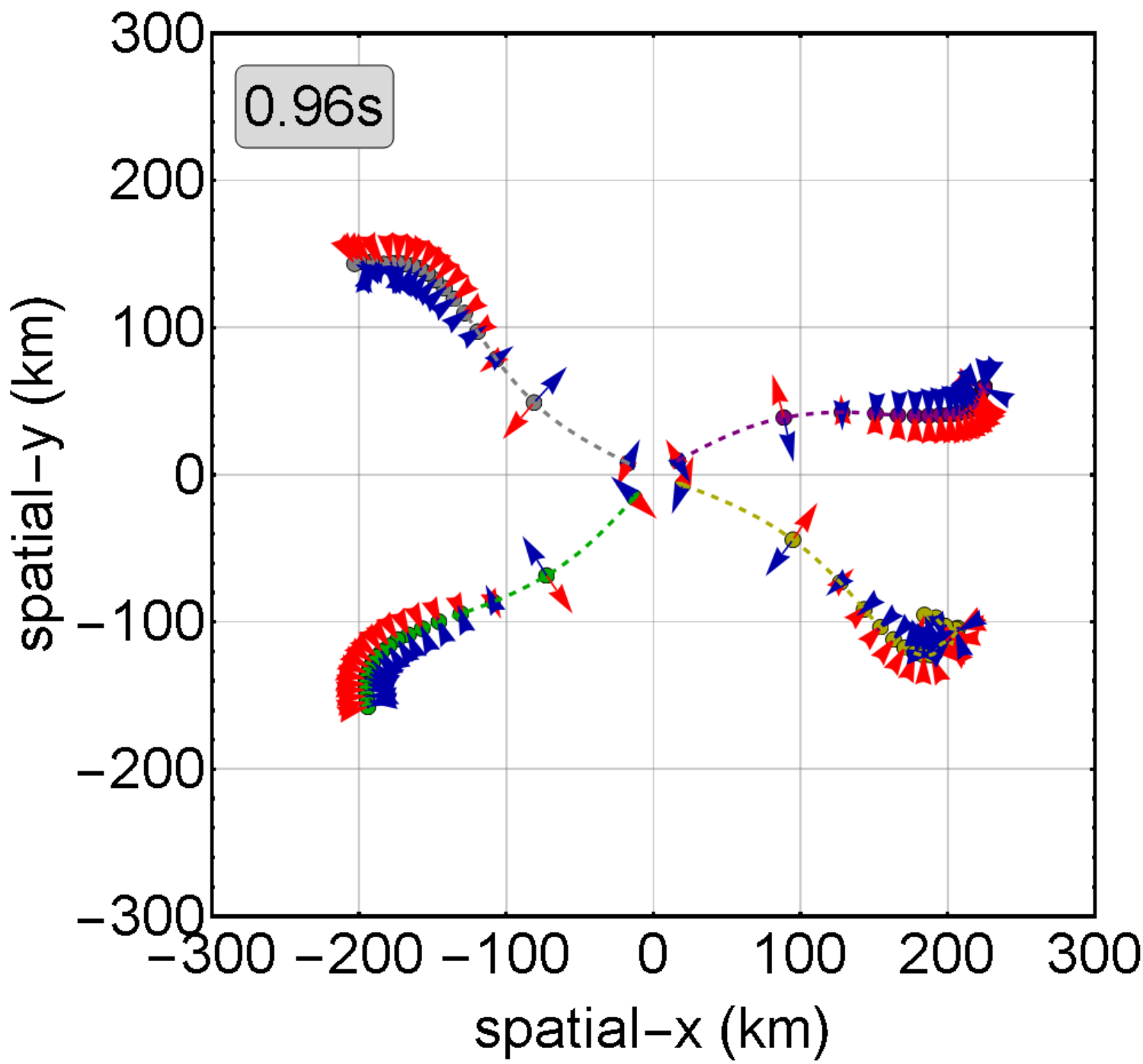}\tabularnewline
\end{tabular}\caption{Motion of the selected ion particles at various snapshots when $\Gamma\simeq0.03$.
The parameters are set the same as in Fig. \ref{fig:Evolution-of-the}.
The red and blue arrows in each graph indicate the directions of the
electric and magnetic (Lorentz) force received by the particles. \label{fig:Motion-of-the-1}}
\par\end{centering}
\end{figure}

The second pattern corresponds to the condition where $\Gamma\simeq0.03$
as in Fig. \ref{fig:Evolution-of-the} and \ref{fig:Motion-of-the-1}.
A prominent difference in this condition is a lack of the breathing
behavior and a flute mode at the later time (Fig. \ref{fig:Evolution-of-the}).
In Fig. \ref{fig:Motion-of-the-1} we have selected four ions and
plotted their relevant electric and magnetic force. The ions move
in an almost straight line. Initially, the Lorentz force is a little
larger than the balancing electric force, hence the particles tend
to curve slight clockwise. At some later time, the electric force
dominates and the ions begin to bend counter-clockwise. 

\begin{figure}

\begin{centering}
\begin{tabular}{cc}
\includegraphics[scale=0.38]{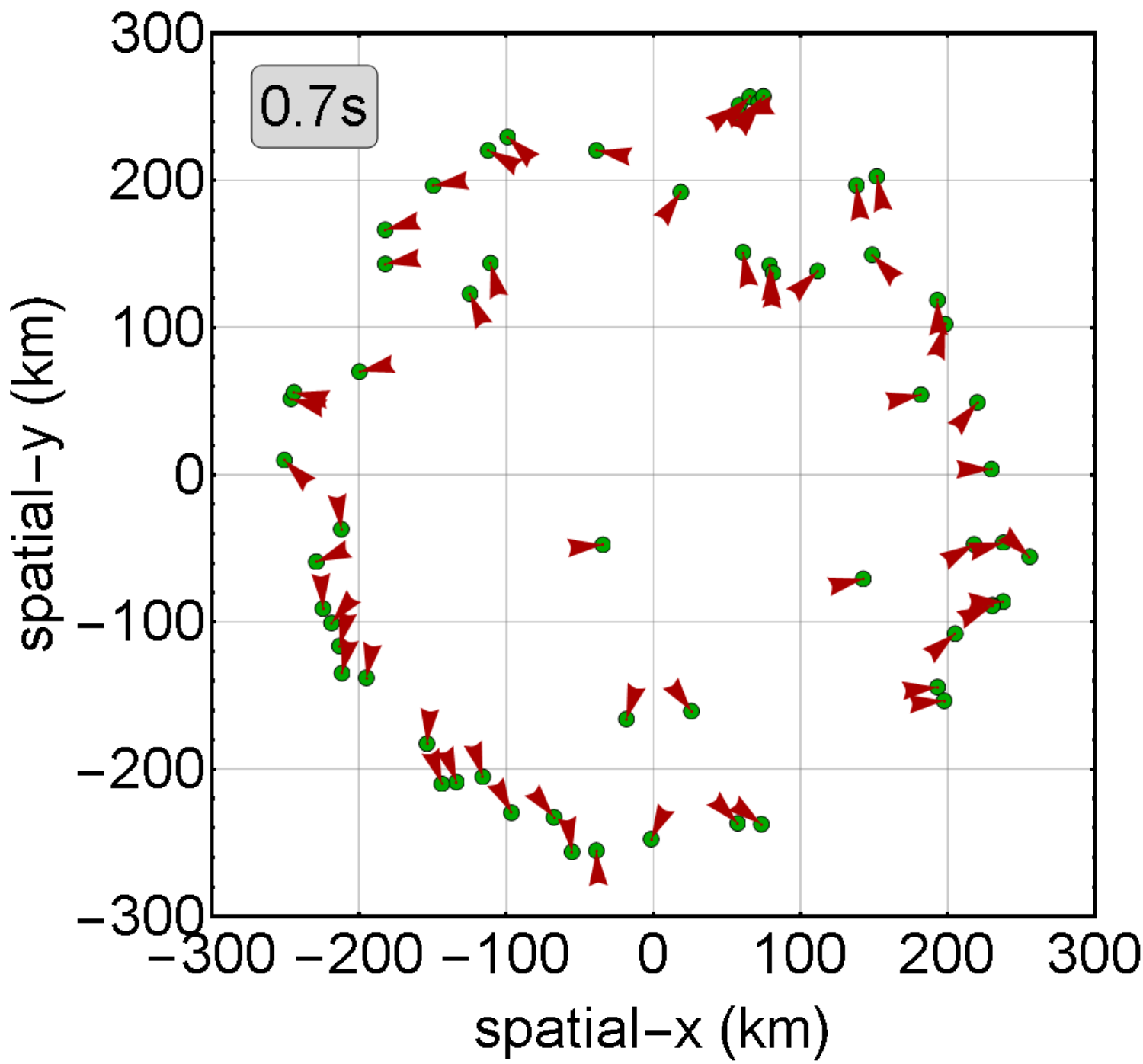} & \includegraphics[scale=0.35]{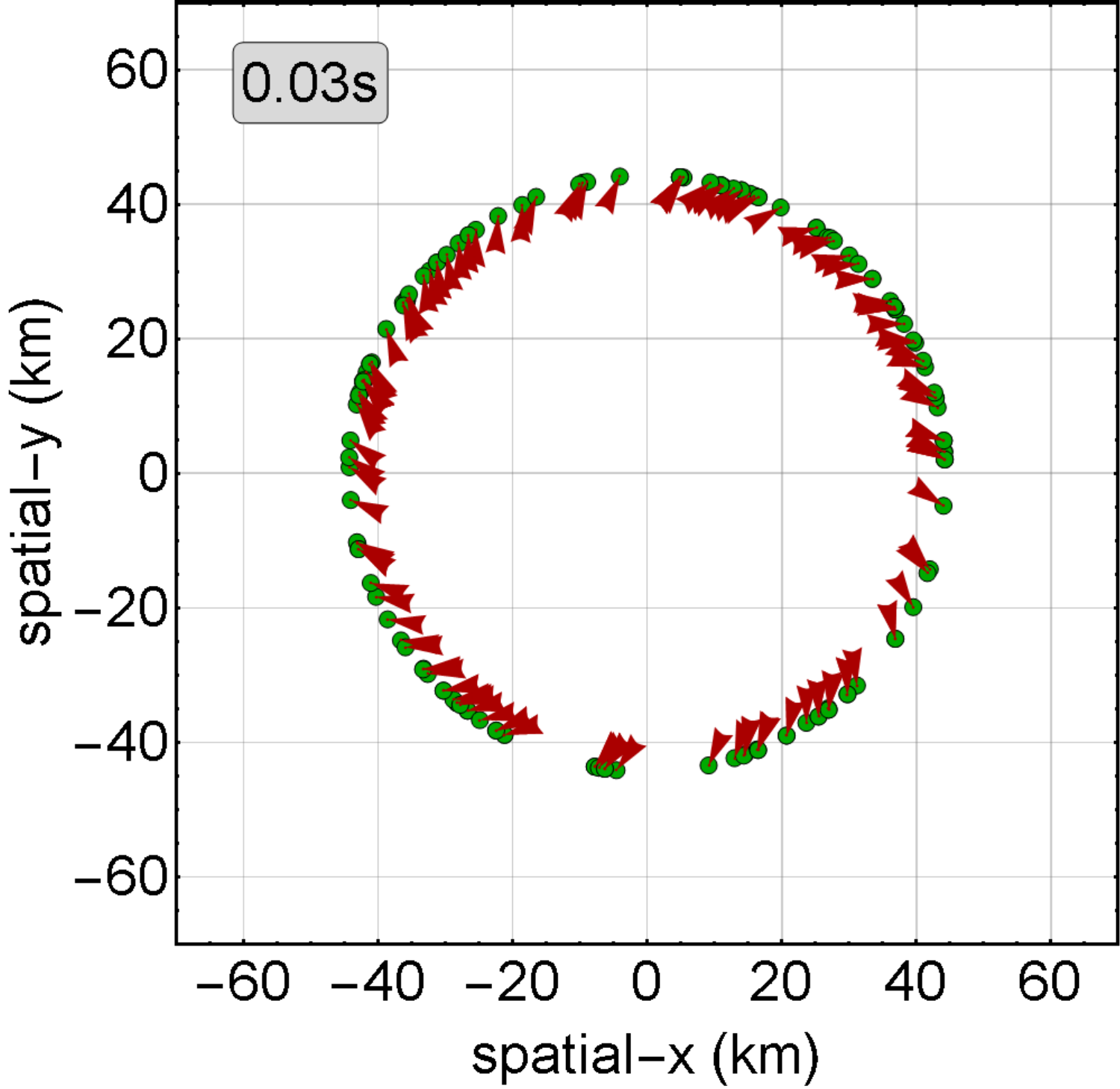}\tabularnewline
\end{tabular}\caption{Directions of the electron drift when $\Gamma\ll1$(left panel) and
$\Gamma\gtrsim1$(right panel). Green dots and red arrows represent
the selected debris ions and the direction of the associated electron
fluid. The arrows are obtained via Eq. (\ref{eq:E in B and U_e}).
The parameters are the same as in Fig. \ref{fig:Motion-of-the} and
\ref{fig:Motion-of-the-1}. We see that the electrons can flow in
different directions at the rim of the expanding plasma depending
on the initial mass of the debris ions.\label{fig:Directions-of-the}}
\par\end{centering}
\end{figure}

These two different behaviors reveal the complexities of the plasma
evolution. A little imbalance of the electric and magnetic force is
able to produce two completely different patterns. To see this, we
recall that in the hybrid model the electric field is perpendicular
to the electron fluid velocity and the magnetic field, i.e., $\mathbf{E}=-\mathbf{u}_{e}\times\mathbf{B}$
(see Eq. (\ref{eq:E in B and U_e})). This allows us to analyze the
drift of the electrons from the depicted electric force in Fig. \ref{fig:Motion-of-the}
and \ref{fig:Motion-of-the-1}. In the breathing pattern where $\Gamma\gtrsim1$,
the electrons initially drift outward. When the electron flow reaches
the out most rim of the torus, it gradually turns \textit{clockwise}.
However, for cases where $\Gamma\ll1$, the electrons drift \textit{counter-clockwise}
at the later time. Therefore, the electrons drift in different directions
at the rim of the expanding plasma depending on the initial mass of
the debris ions (Fig. \ref{fig:Directions-of-the}).

\begin{figure}
\begin{centering}
\begin{tabular}{cc}
\includegraphics[scale=0.38]{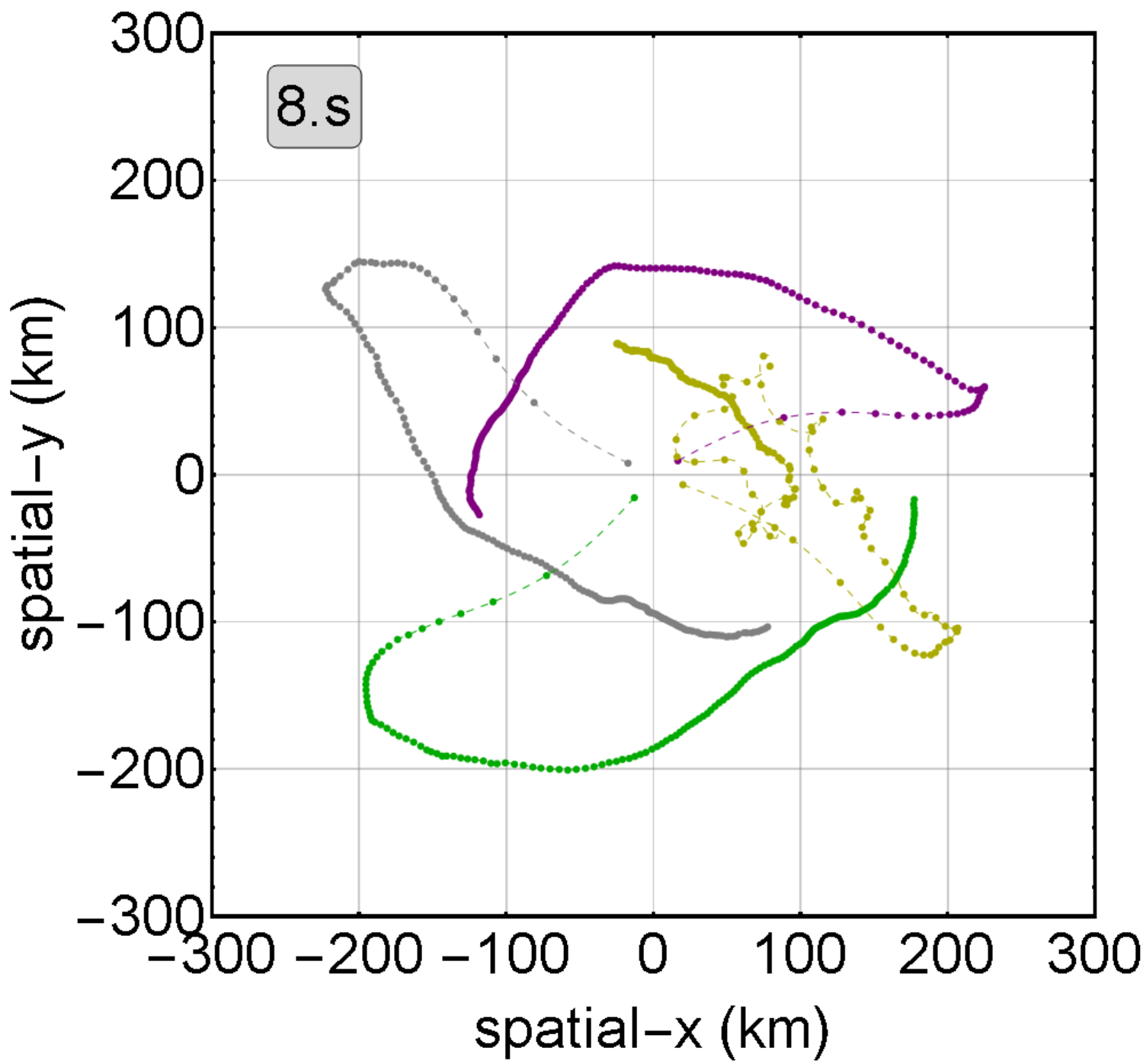} & \includegraphics[scale=0.34]{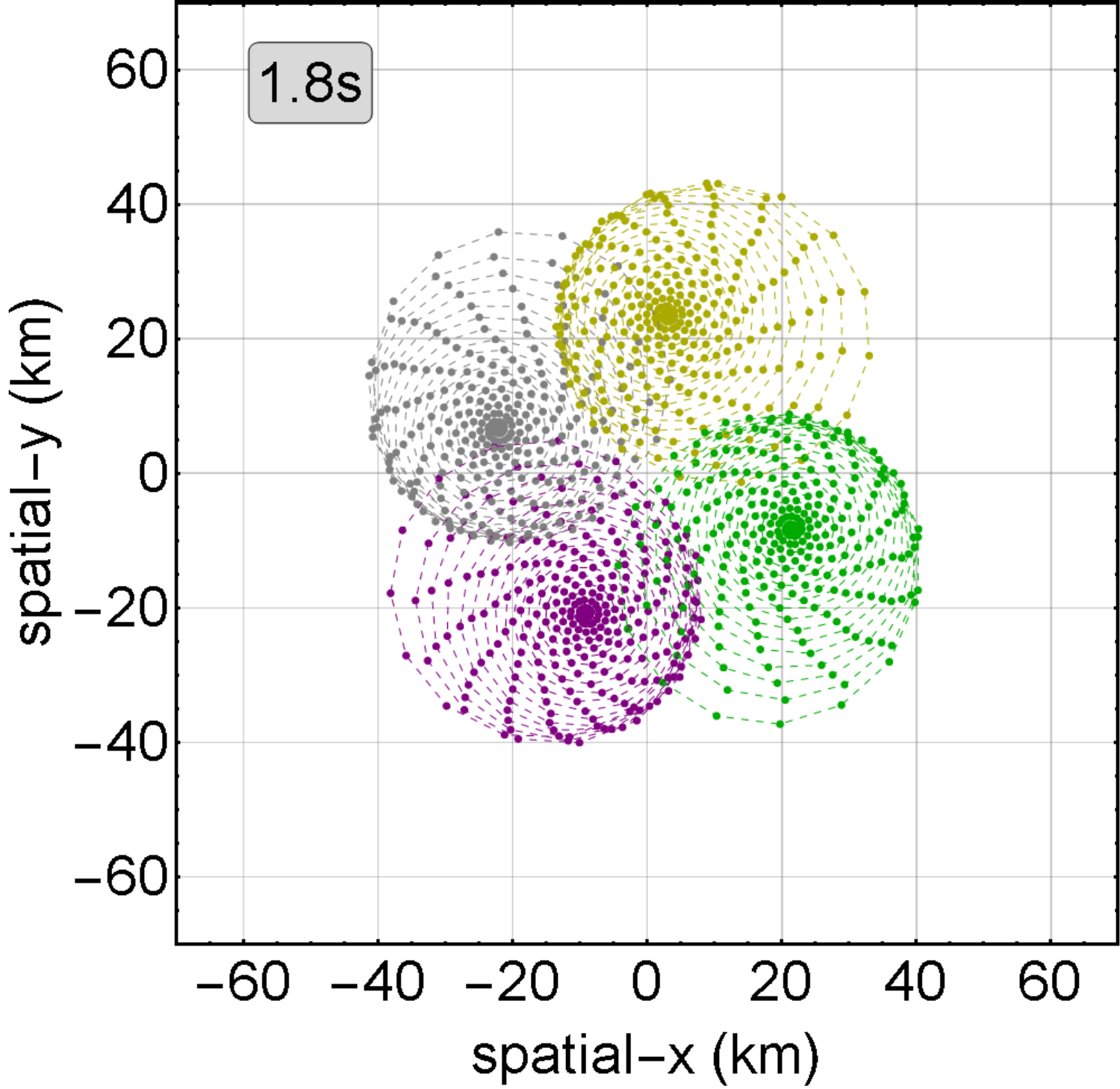}\tabularnewline
\end{tabular}\caption{Trajectories of the heavy ions when $\Gamma\ll1$(left panel) and
$\Gamma\gtrsim1$(right panel). The time stamp on the left corner
of each graph denotes that only the trajectories before this time
is tracked. All parameters are the same as in Fig. \ref{fig:Motion-of-the}
and \ref{fig:Motion-of-the-1}. \label{fig:Trajectories-of-the}}
\par\end{centering}
\end{figure}

Now we analyze the stopping behavior of the two conditions. When $\Gamma\ll1$
a strong flute mode will occur, therefore the flute instability should
influence the way the heavy ions behave. In Fig. \ref{fig:Trajectories-of-the}
we also plot the trajectories of the four selected ions. The left
panel of Fig. \ref{fig:Trajectories-of-the} shows that the ions are
turning counter-clockwise at the later time. This is consistent with
the drift direction of the electrons which is also counter-clockwise.
However, the right panel demonstrates a rather different behavior
--- the heavy ions are rotating clockwise in the ambient magnetic
field and gradually lose energy. 

Finally, we may find some clues of these two phenomena from the building-up
equations, i.e., Eq. (\ref{eq:E}) $\sim$ (\ref{eq:neutral condition}).
In Fig. \ref{fig:Motion-of-the} and \ref{fig:Motion-of-the-1}, we
can see that the electric force is mainly in the $\theta$ direction.
In Fig. \ref{fig:Motion-of-the-1}, the balancing Lorentz force is
also in approximately the $\theta$ direction initially. When the
electric force in the $\theta$ direction becomes prominent, it curves
the trajectories of the ions at the later time (see Fig. \ref{fig:Motion-of-the-1}).
The debris ions will then evolve into a flute like pattern. On the
contrary, if the Lorentz force is dominant, the debris ions will circulate
in the ambient geomagnetic field and evolve toward the breathing like
pattern. Therefore, comparing the $\theta$ component of the electric
field will suffice the analysis. 

Given that the difference of the phenomenon starts from an imbalance
of electric and magnetic force acted upon the heavy ions, we rewrite
Eq. (\ref{eq:E}) in terms of the number densities of the ions $n_{\text{D}}$
and the background ions $n_{\text{B}}$,
\begin{eqnarray}
\mathbf{E} & = & -\frac{1}{(n_{\text{D}}+n_{\text{B}})}\left[\left(n_{\text{D}}\mathbf{u}_{\text{D}}+n_{\text{B}}\mathbf{u}_{\text{B}}\right)\times\mathbf{B}-\frac{1}{\mu_{0}}\left(\mathbf{B}\cdot\nabla\right)\mathbf{B}+\frac{1}{2\mu_{0}}\nabla\mathbf{B}^{2}\right],\label{eq:E in n_D and n_B}
\end{eqnarray}
where we have used the identity $\left(\nabla\times\mathbf{B}\times\mathbf{B}\right)=\left(\mathbf{B}\cdot\nabla\right)\mathbf{B}-\frac{1}{2}\nabla\mathbf{B}^{2}$.
All ions have charge state $+1$ with $\mathbf{J}_{\text{D}}=en_{\text{D}}\mathbf{u}_{\text{D}}$
and $\mathbf{J}_{\text{B}}=en_{\text{B}}\mathbf{u}_{\text{B}}$. 

From Eq. (\ref{eq:A4}) in the Appendix, we have
\begin{eqnarray}
E_{\theta} & \simeq & -\frac{1}{(n_{\text{D}}+n_{\text{B}})}\left[-\left(n_{\text{D}}u_{\text{D},r}+n_{\text{B}}u_{\text{B},r}\right)B_{z}-B_{z}\frac{\partial B_{\theta}}{\partial z}\right].\label{eq:E_theta}
\end{eqnarray}
 For a specific heavy ion with radial velocity $v_{\text{D},r}$,
the ratio of the electric (in the $\theta$ direction) to Lorentz
force has the following form
\begin{eqnarray}
\frac{E_{\theta}}{v_{\text{D},r}B_{z}} & = & \frac{n_{\text{D}}u_{\text{D},r}+n_{\text{B}}u_{\text{B},r}+\partial B_{\theta}/\partial z}{(n_{\text{D}}+n_{\text{B}})v_{\text{D},r}},\label{eq:E_theta/B}
\end{eqnarray}
where $v_{\text{D},r}B_{z}$ gives the Lorentz force in the $\theta$
direction. For cases where $\Gamma\ll1$ as in Fig. \ref{fig:Motion-of-the-1},
the electric force will exceed the Lorentz force at some later time,
i.e., $E_{\theta}>v_{\text{D},r}B_{z}$. Hence
\begin{eqnarray}
\frac{n_{\text{D}}u_{\text{D},r}+n_{\text{B}}u_{\text{B},r}+\partial B_{\theta}/\partial z}{(n_{\text{D}}+n_{\text{B}})v_{\text{D},r}} & > & 1.\label{eq:u_b u_d}
\end{eqnarray}
Eq. (\ref{eq:u_b u_d}) gives a rough criteria for the breathing mode.
One sees that the flute like pattern does not purely rely on the number
densities, rather it depends on both the number densities and the
structure of the magnetic field. However, we emphasize that the criteria
in Eq. (\ref{eq:u_b u_d}) is directly concluded from the numerical
results of the hybrid model. The real plasma evolution should in principle
contain the high frequency instabilities and the charge separation
effect, etc. 

\section{Conclusion\label{sec:Summary-and-conclusion}}

In the present work, we have studied the phenomenon where a patch
of heavy ions expand in a cold background plasma with the existence
of an ambient magnetic field. We have simulated two conditions where
the initial total mass of the heavy ions is different. The hybrid
model results of these two cases demonstrate a distinct collective
behavior of the ion's motion. Due to the imbalance of electric and
magnetic force, the heavy ions may involve into a flute like pattern
or a breathing like pattern. These patterns reflect a different stopping
and collective behavior of the ions. Finally, we also give a rough
criteria for the flute like pattern to occur.

Since the contribution of charge separation may not be negligible,
a full PIC (Particle-In-Cell) treatment\cite{Bowers2008} or the direct
solving of the coupled relativistic Boltzmann-Maxwell's equations\cite{Zhang2020}
are required. The hybrid simulation model used in this paper assumes
that the electrons always appear in such a way that the ions are neutralized,
and meanwhile, we have also assumed that displacement current can
be neglected in the formulation. These assumptions may not be truly
satisfied and may influence the evolution of the plasma system. We
will study these factors in detail in our future work. 

\begin{acknowledgments}
The authors are thankful for the technical support from Prof. Zhong-Qi Wang in  Beijing Institute of Technology. The work is supported by the National Key Research and Development Program of China (NKRDPC) under the grant number:2020YFA0709800. The data that support the findings of this study are openly available in Harvard Dataverse at https://doi.org/10.7910/DVN/MLZKPS.
\end{acknowledgments}

\section{Appendix\label{sec:Appendix}}

In cylindrical coordinate the following relations hold
\begin{eqnarray}
\mathbf{u}\times\mathbf{B} & = & \left(u_{\theta}B_{z}-u_{z}B_{\theta}\right)\hat{r}\nonumber \\
 &  & +\left(u_{z}B_{r}-u_{r}B_{z}\right)\hat{\theta}\nonumber \\
 &  & +\left(u_{r}B_{\theta}-u_{\theta}B_{r}\right)\hat{z},\label{eq:A1}
\end{eqnarray}
\begin{eqnarray}
\frac{1}{2\mu_{0}}\nabla\mathbf{B}^{2} & = & \frac{1}{2\mu_{0}}\left(\frac{\partial\mathbf{B}^{2}}{\partial r}\hat{r}+\frac{1}{r}\frac{\partial\mathbf{B}^{2}}{\partial\theta}\hat{\theta}+\frac{\partial\mathbf{B}^{2}}{\partial z}\hat{z}\right),\label{eq:A2}
\end{eqnarray}
and
\begin{eqnarray}
-\frac{1}{\mu_{0}}\left(\mathbf{B}\cdot\nabla\right)\mathbf{B} & = & -\frac{1}{\mu_{0}}\left[\left(B_{r}\frac{\partial B_{r}}{\partial r}+\frac{B_{\theta}}{r}\frac{\partial B_{r}}{\partial\theta}+B_{z}\frac{\partial B_{r}}{\partial z}-\frac{B_{\theta}^{2}}{r}\right)\hat{r}\right.\nonumber \\
 &  & +\left(B_{r}\frac{\partial B_{\theta}}{\partial r}+\frac{B_{\theta}}{r}\frac{\partial B_{\theta}}{\partial\theta}+B_{z}\frac{\partial B_{\theta}}{\partial z}+\frac{B_{\theta}B_{r}}{r}\right)\hat{\theta}\nonumber \\
 &  & \left.+\left(B_{r}\frac{\partial B_{z}}{\partial r}+\frac{B_{\theta}}{r}\frac{\partial B_{z}}{\partial\theta}+B_{z}\frac{\partial B_{z}}{\partial z}\right)\hat{z}\right],\label{eq:A3}
\end{eqnarray}
where $\mathbf{u}\equiv n_{\text{D}}\mathbf{u}_{\text{D}}+n_{\text{B}}\mathbf{u}_{\text{B}}$.
If we restrict our analysis in the spatial $x-y$ plane, the $\hat{r}$
and $\hat{\theta}$ components of the electric field $E_{r}$ and
$E_{\theta}$ will suffice. In the expansion of the debris ions, $u_{r}$
should be larger than $u_{z}$, hence $u_{z}B_{r}\ll u_{r}B_{z}$
in Eq. (\ref{eq:A1}). 

\begin{figure}

\begin{centering}
\begin{tabular}{ccc}
\includegraphics[scale=0.23]{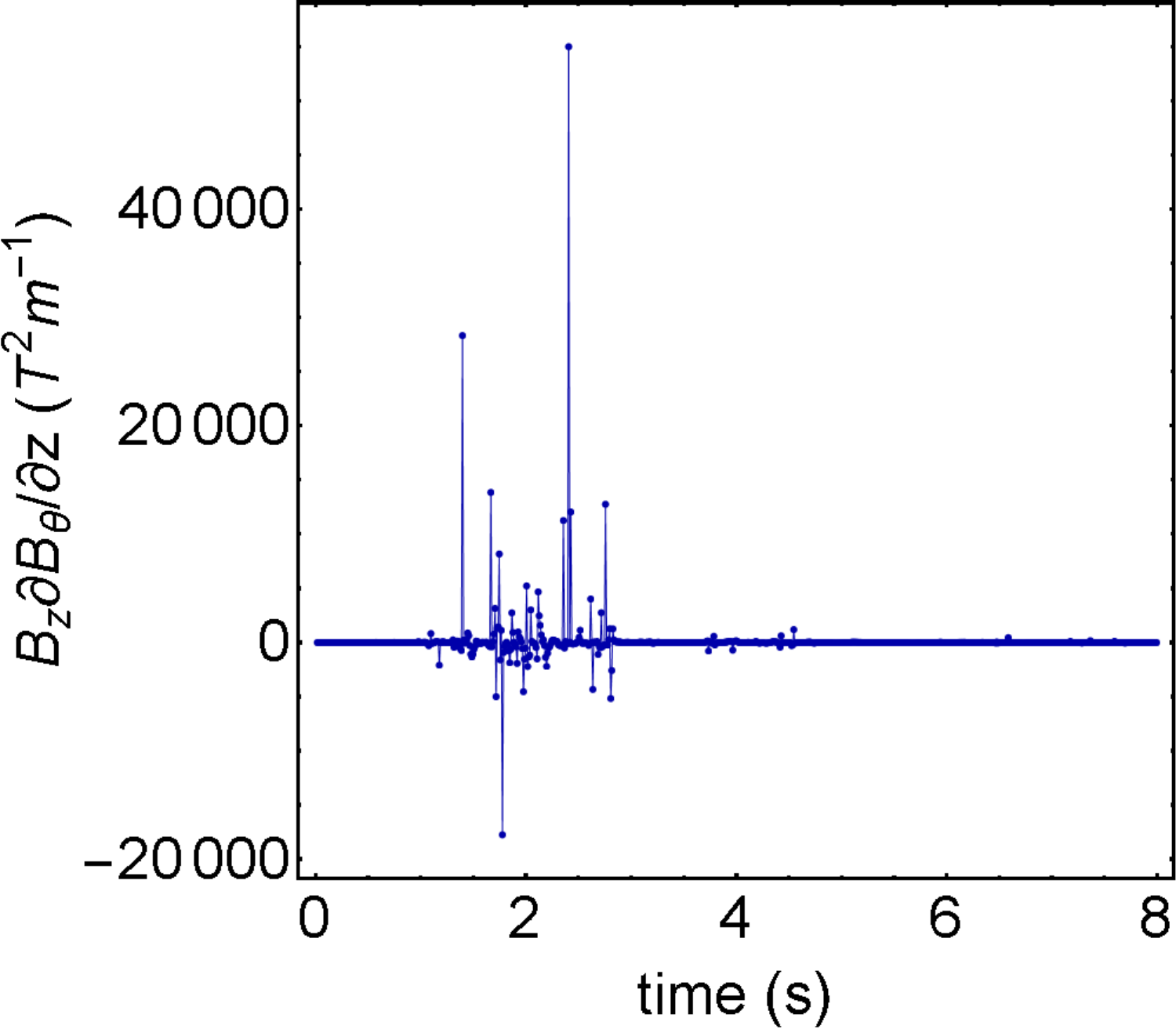} & \includegraphics[scale=0.24]{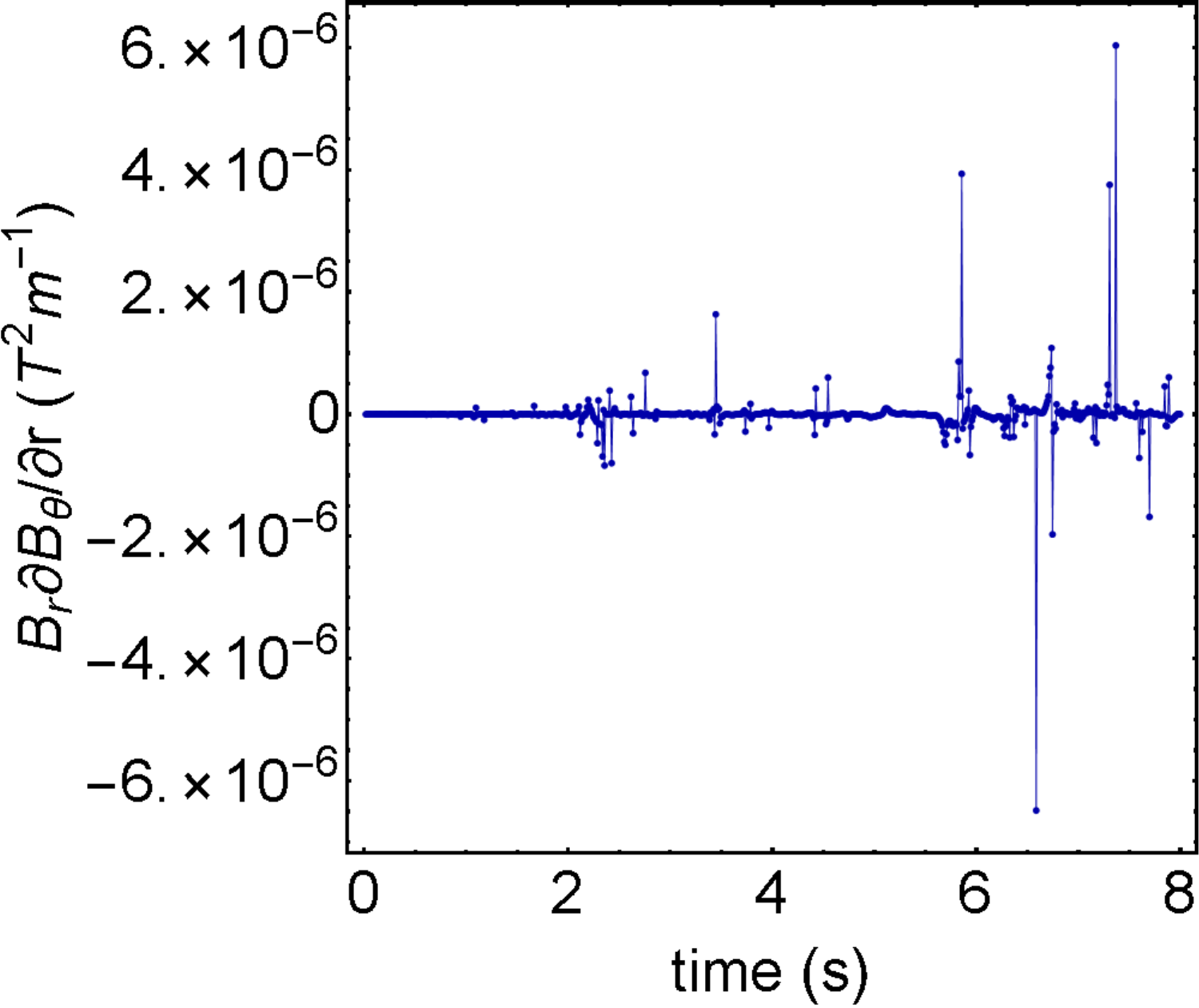} & \includegraphics[scale=0.25]{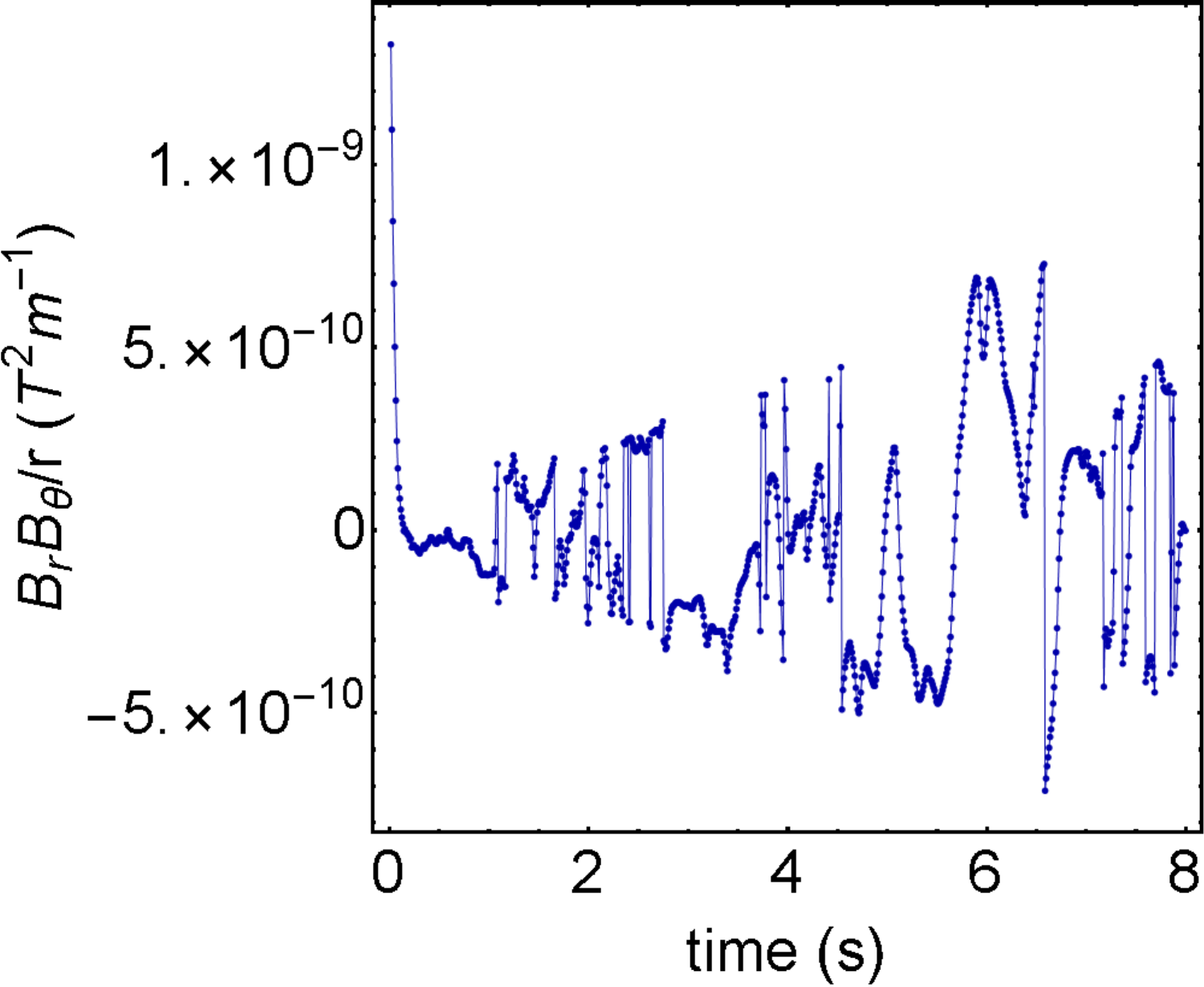}\tabularnewline
\end{tabular}
\par\end{centering}
\caption{Comparison of the terms $B_{z}\frac{\partial B_{\theta}}{\partial z}$,
$B_{r}\frac{\partial B_{\theta}}{\partial r}$ and $\frac{B_{\theta}B_{r}}{r}$.
The parameters are the same as in Fig. \ref{fig:Evolution-of-the}.
The data is extracted from one of the four selected particles in Fig.
\ref{fig:Motion-of-the-1}. We can see that the contribution of the
term $B_{z}\frac{\partial B_{\theta}}{\partial z}$ is significantly
larger than that of other terms.\label{fig:Comparison-of-the}}

\end{figure}

From Fig. \ref{fig:Motion-of-the} and \ref{fig:Motion-of-the-1},
we can see that the electric force is nearly symmetrical in the $\theta$
direction. This allows us to drop the terms $\frac{1}{r}\frac{\partial\mathbf{B}^{2}}{\partial\theta}\hat{\theta}$
and $\frac{B_{\theta}}{r}\frac{\partial B_{\theta}}{\partial\theta}$.
Meanwhile, the contribution of the term $B_{z}\frac{\partial B_{\theta}}{\partial z}$
is larger than terms $B_{r}\frac{\partial B_{\theta}}{\partial r}$
and $\frac{B_{\theta}B_{r}}{r}$ (see the numerical comparison in
Fig. \ref{fig:Comparison-of-the}). Therefore, we can express the
$\hat{\theta}$ component of the electric field $E_{\theta}$ in terms
of $B_{z}$, 
\begin{eqnarray}
E_{\theta} & \simeq & -\frac{1}{(n_{\text{D}}+n_{\text{B}})}\left[-\left(n_{\text{D}}u_{\text{D},r}+n_{\text{B}}u_{\text{B},r}\right)B_{z}-B_{z}\frac{\partial B_{\theta}}{\partial z}\right].\label{eq:A4}
\end{eqnarray}
One may find that Eq. (\ref{eq:A4}) is different from the one expressed
in \cite{HEWETT2011}, where
\begin{eqnarray}
E_{\theta} & \simeq & -\frac{1}{(n_{\text{D}}+n_{\text{B}})}\left[-\left(n_{\text{D}}u_{\text{D},r}+n_{\text{B}}u_{\text{B},r}\right)B_{z}\right].\label{eq:A5}
\end{eqnarray}
This is because in our analysis, the deviation of two different phenomena
takes place at a much later time than in \cite{HEWETT2011}. Thus,
the contribution of the term $\nabla\times\mathbf{B}\times\mathbf{B}$
cannot be neglected.

\nocite{*}
\bibliography{Ref}

\end{document}